\documentclass[letterpaper]{article}

\usepackage[utf8]{inputenc}
\usepackage{float}
\usepackage{arxiv}
\usepackage{enumitem}
\usepackage[english]{babel}
\usepackage{amssymb}
\usepackage{amsmath,mathtools}
\usepackage{xcolor}
\usepackage{graphicx}
\usepackage{multirow}
\usepackage{overpic}
\usepackage{psfrag}
\usepackage{bm}
\usepackage{url}
\usepackage{epsfig}
\usepackage{marvosym}
\usepackage{verbatim}
\usepackage{rotating}
\usepackage{subcaption}
\usepackage{caption}
\usepackage{hyperref}
\hypersetup{colorlinks=true,linkcolor=red,citecolor=blue}
\usepackage{ulem}
\usepackage{lmodern}
\usepackage{booktabs} 
\usepackage{moresize} 
\usepackage{siunitx}
\usepackage{lineno}
\usepackage{lipsum}
\usepackage{tikz}
\usepackage{pgfplots}
\usepgfplotslibrary{colormaps} 
\usetikzlibrary{pgfplots.colormaps} 
\pgfplotsset{compat=1.14}
\usepgfplotslibrary{fillbetween}
\usepgfplotslibrary{statistics}
\usetikzlibrary{shapes, arrows, arrows.meta, calc, positioning, decorations,
decorations.pathmorphing, decorations.text, decorations.markings, fit, plotmarks}
\pgfplotsset{translate gnuplot=true}
\usepackage{etex} 								
\usepgfplotslibrary{external} 						
\tikzsetexternalprefix{ext/}							

\definecolor{myblue1}	{RGB}{0,177,234}				
\definecolor{myblue2}	{RGB}{76,200,239}				
\definecolor{myblue3}	{RGB}{127,215,244}				
\definecolor{myblue4}	{RGB}{178,231,248}				
\definecolor{myblue5}	{RGB}{198,251,255}				
\definecolor{mybluegray1}{RGB}{0,127,167}				
\definecolor{mybluegray2}{RGB}{76,165,193}				
\definecolor{mybluegray3}{RGB}{127,191,211}				
\definecolor{mybluegray4}{RGB}{178,216,228}				
\definecolor{mygray1}	{RGB}{76,84,93}				
\definecolor{mygray2}	{RGB}{129,135,141}				
\definecolor{mygray3}	{RGB}{165,169,174}				
\definecolor{mygray4}	{RGB}{201,203,206}				
\definecolor{myorange1}	{RGB}{255,126,46}				
\definecolor{myorange2}	{RGB}{255,164,108}				
\definecolor{myorange3}	{RGB}{255,190,150}				
\definecolor{myorange4}	{RGB}{255,216,192}				
\definecolor{mypurple1}{RGB}{89,89,171}
\definecolor{mypurple4}{RGB}{189,189,231}

\newcommand\red[1]{\textcolor{black}{#1}}

\pgfplotsset{
    colormap={custom_map}{[5pt]
            rgb255(0pt)=(255,126,46);
            rgb255(500pt)=(255,190,150);
            rgb255(1000pt)=(0,177,234);
            rgb255(1500pt)=(127,215,244);
    },
}

\newcommand\ie{\textit{i.e.}\,\,}

\renewcommand{\emph}[1]{\textit{#1}}
\setenumerate{label=\textcolor{mybluegray1}{$\bm{\diamond}$},itemsep=0.0pt,topsep=5pt}

\newcommand{\mut}{{\mu_{t}}}
\newcommand{\nut}{\tilde{\nu}}
\newcommand{\uu}{{\boldsymbol u}}

\pgfplotsset{
	colormap/rdbur/.style={
		colormap={rdbur}{
			rgb255(0cm)=(20,46,97); 
			rgb255(1cm)=(52,100,171); 
			rgb255(2cm)=(83,146,194); 
			rgb255(3cm)=(153,197,222); 
			rgb255(4cm)=(211,229,240);
			rgb255(5cm)=(247,247,247);
			rgb255(6cm)=(250,219,200);
			rgb255(7cm)=(238,165,132);
			rgb255(8cm)=(207,94,80);
			rgb255(9cm)=(171,10,46);
			rgb255(10cm)=(99,0,32);
			}
		}
	}

\newcommand*{\shifttext}[2]{%
  \settowidth{\@tempdima}{#2}%
  \makebox[\@tempdima]{\hspace*{#1}#2}%
}

\title{Robust Deep Learning For Emulating Turbulent Viscosities}

\def\size{7.2cm}
\author{
	\parbox{\size}{\centering Aakash Patil}\\
	MINES ParisTech, CEMEF\\
	PSL - Research University\\
	06904 Sophia Antipolis, France\\
	\texttt{aakash.patil@mines-paristech.fr}
\And
	\parbox{\size}{\centering Jonathan Viquerat\thanks{Corresponding author}}\\
	MINES ParisTech, CEMEF\\
	PSL - Research University\\
	06904 Sophia Antipolis, France\\
	\texttt{jonathan.viquerat@mines-paristech.fr}
\And
	\parbox{\size}{\centering  Aurélien Larcher}\\
	MINES ParisTech, CEMEF\\
	PSL - Research University\\
	06904 Sophia Antipolis, France\\
	\texttt{aurelien.larcher@mines-paristech.fr}
\And
	\parbox{\size}{\centering George El Haber}\\
	MINES ParisTech, CEMEF\\
	PSL - Research University\\
	06904 Sophia Antipolis, France\\
	\texttt{georges.el\_haber@mines-paristech.fr}
\And
	\parbox{\size}{\centering Elie Hachem}\\
	MINES ParisTech, CEMEF\\
	PSL - Research University\\
	06904 Sophia Antipolis, France\\
	\texttt{elie.hachem@mines-paristech.fr}
}

\begin{document}
\newgeometry{left=3cm,right=3cm,top=3cm,bottom=3cm}
\maketitle

\begin{abstract}
From the simplest models to complex deep neural networks, modeling turbulence with machine learning techniques still offers multiple challenges. In this context, the present contribution proposes a robust strategy using patch-based training to learn turbulent viscosity from flow velocities, and demonstrates its efficient use on the Spalart-Allmaras turbulence model. Training datasets are generated for flow past two-dimensional obstacles at high Reynolds numbers and used to train an auto-encoder type convolutional neural network with local patch inputs. Compared to a standard training technique, patch-based learning not only yields increased accuracy but also reduces the computational cost required for training.
\end{abstract}


\section{Introduction}

Computational fluid dynamics (CFD) is an essential asset for research and industrial applications. Despite advances in computational power over the years, industrial CFD tools still largely rely on the Reynolds Averaged Navier-Stokes (RANS) turbulence models due to cost-savings and lesser time-to-solution offered by RANS when compared with intensive Large Eddy Simulations (LES) and Direct Numerical Simulations (DNS), especially for flows at high-Reynolds numbers. Among the variety of one-equation to many equations RANS models, the Spalart Allmaras (SA) turbulence model \cite{spalart1992one}, that solves for the kinematic eddy turbulent viscosity, \red{and has not been derived from the existing turbulent kinetic energy-based RANS models. In this sense, it can be said as a proper one-equation model which does not require knowledge of a specific problem and additional advantages include numerical stability as well as reliability for convergence of results. Due to these reasons, the SA model} has been widely used, documented, and serves as a benchmark turbulence model for many CFD applications \cite{ferziger2002computational, spalart2000strategies}. 

During the past decade, the coupling of CFD algorithms with deep learning methods, and especially neural networks, have progressed rapidly. \red{Such couplings offer new perspectives and opportunities to assist the existing CFD solvers, by leveraging the power of deep learning and provide an additional probe into our understanding of the modelling of turbulent flows. Industrial applications of such couplings include increase efficiency and accuracy for the simulation of complex flows, but also faster design-cycles as well as empowered real-time digital twins.} One of the early works on the use of neural networks in fluid dynamics was reported in \cite{milano2002neural}, where near-wall velocity fields were predicted by comparing the equivalency of prediction from neural networks with proper orthogonal decomposition-based reconstruction. Several works, such as  \cite{Yarlanki2012, CHEUNG2011, Kato2014}, have used velocity field data to predict model parameters and their probability distributions to quantify and reduce modeling errors. A proper framework for using machine learning methods in the area of fluid dynamics was laid down since the works of \cite{PARISH2016758} and \cite{xiao2016quantifying} in which the authors have demonstrated the use of these methods for turbulence modeling in the form of estimation of model uncertainties using machine learning. Overall, a paradigm for data-driven predictive modeling of turbulent flows by systematic implementation of machine learning and inverse modeling was described in \cite{wang2017physics, singh2017data, singh2017machine, singh2016using, singh2017augmentation}. Moreover, the direct prediction of Reynolds stresses for RANS and prediction of deconvoluted direct numerical simulation have been proposed in \cite{ling2015evaluation, ling2016reynolds, vollant2017subgrid, maulik2017neural}. Similar works involving re-generating turbulence statistics as well as super-resolution have been demonstrated in \cite{fukami2019synthetic,mohan2019compressed,beck2019deep,kim2019deep,fukami2019super,fukami2020machine} whereas \cite{zhao2019turbulence,taghizadeh2020turbulence} investigated the coupling of RANS with machine learning optimization. Deep learning has also been utilized in CFD for a variety of related tasks such as drag prediction as described in \cite{viquerat2020supervised} and flow-reconstruction along with uncertainty estimation in \cite{chen2021twin}. The use of machine learning in the turbulence modeling community has been summarized in the recent reviews by Duraisamy \textit{et al.} \cite{duraisamy2019turbulence} and Zang \textit{et al.} \cite{zhang2019recent}.

The Spalart-Allmaras turbulence model has also been subject to machine learning-based investigations in several works. In 2015, Tracey \textit{et al.} \cite{tracey2015machine} demonstrated one of the first works on the SA model using neural networks to predict a part of the RANS closure model (namely the source terms of the eddy viscosity transport). Their study features hand-picking of features from the SA eddy viscosity transport equation in order to predict its source term. Later on, Singh \textit{et al.} \cite{singh2017machine} followed a similar procedure by exploiting hand-picking of input features deploying neural networks for the prediction of source terms of SA eddy viscosity transport, and later presented both a priori and a posteriori analysis. More recently,  \cite{liang2019developing,maulik2020turbulent} proposed to predict the eddy viscosity from input features consisting of data from Navier-Stokes and transport equations, while \cite{pal2020deep} used neural networks for predicting subgrid-scale viscosity in the geophysical applications. 

Although convolutional neural networks are traditionally trained using full-scale inputs, patch-wise deep learning models have been successfully applied in the past in the computer-vision community. In particular, Long \textit{et al.} proposed a work on object detection \cite{long2015fully} where data was spatially divided into patches of information, which were then fed to the model with a primary intention of reducing memory consumption during training. This study proved efficient not only in terms of reductions in error but also in reduced memory consumption during the training. Also, it has been demonstrated in the image classification tasks in \cite{farabet2012learning, pinheiro2014recurrent} that the patchwise training can correct class imbalance as well as assist in the spatial correlation of dense patches. More than saving training memory, patch-wise training offers a great opportunity for deep learning research in CFD, primarily because the spatial data can be divided into small patches of neighboring nodes to achieve global as well as local learning, independent of the size of the domain. It also offers a way for CFD data augmentation by increasing the quantity of the same data with flipping and rotation of patches. 

In the present contribution, we aim at learning the SA turbulent viscosity from the velocity field using a convolutional neural network trained following a patch-based approach. The proposed novelties are (i) the blind-learning of SA eddy viscosity from input velocities, without any hand-picking, manual feature selection, non-dimensionalization, or tailored losses, and (ii) a novel patch-based learning strategy with an auto-encoder type convolutional neural network, provided as a way towards generalized deep learning in turbulence modeling. The remaining of the paper is organized as follows: first, the problem setup and its governing equations are described, and the methods used for the dataset generation are covered; then, the selected network architecture is presented, and the patch-based learning procedure is described thoroughly; finally, the interests of the proposed approach are assessed, and results are discussed and compared to baseline solutions.

\section{Problem setup \& data generation}

\subsection{Governing equations}
\label{section:equations}

The evolution of the velocity $\boldsymbol{u}$ and pressure $p$ in an incompressible fluid flow with given positive constant density $\rho$  and dynamic viscosity $\mu$ is governed by the Navier-Stokes equations:

\begin{equation}
\label{eq:ns_equation1}
\left\{
\begin{array}{l}
\rho\;(\partial_{t} \boldsymbol{u} + \boldsymbol{u} \cdot \mathbf{\nabla} \boldsymbol{u}) - \mathbf{\nabla} \cdot \boldsymbol{\sigma}  = \boldsymbol{f},\\[1ex]
\mathbf{\nabla} \cdot \boldsymbol{u} = 0,
\end{array}
\right.
\end{equation}

where $\boldsymbol{\sigma} =2 \mu\; \boldsymbol{\varepsilon}(\boldsymbol{u})- p\;{\mathbf{I}_d}$ is the Cauchy stress tensor for a Newtonian fluid, $\boldsymbol{\varepsilon}(\boldsymbol{u})$ the strain-rate tensor, and ${\mathbf{I}_d}$ the $d$-dimensional identity tensor. Equations \eqref{eq:ns_equation1} are supplemented with adequate boundary and initial conditions, to be specified. Reynolds-Averaged Navier-Stokes (RANS) equations are then obtained by applying the Reynolds decomposition to the system \eqref{eq:ns_equation1}, such that velocity and pressure are expressed as the sum of a mean-field and a fluctuation. Applying a time averaging operator to the resulting expressions yields a forcing term under the form of the divergence of the so-called Reynolds stress tensor. The latter consists of correlations of velocity fluctuations and accounts for the effect of the turbulent fluctuations on the averaged flow. In the Boussinesq approximation, first-order closure of the system of averaged equations amounts to a mean gradient hypothesis: turbulence is therefore modelled as an additional diffusivity called eddy viscosity $\mut$. The eddy viscosity $\mut$ itself proceeds from a model involving one or more turbulent scales, each of which is the solution of a nonlinear convection-diffusion-reaction equation. For additional details, the reader is referred to the works of \cite{pope2001turbulent} on turbulent flows.

The turbulence model chosen to compute the eddy viscosity is the one-equation Spalart-Allmaras (SA) model \cite{spalart1992one}, which describes the evolution of the kinematic eddy viscosity by solving a convection-diffusion-reaction problem and serves as baseline for future testing of other models. Applying this model, the eddy viscosity $\mut$ in the Navier-Stokes equations is obtained by $\mut = \rho\,\nut f_{v1}$, where $f_{v1}$ is a given damping function to enforce linear profile in the viscous sublayer. The turbulent scale $\nut$ is itself governed by the following nonlinear convection-diffusion-reaction equation: 

\begin{equation}
\label{eq:SA_eq}
\frac{\partial \nut}{\partial t} + \uu \cdot \nabla \nut - c_{b1}(1-f_{t2})\tilde{S}\nut 
	+ \left[c_{w1}f_w - \frac{c_{b1}}{\kappa^2}f_{t2}\right]\left(\frac{\nut}{d}\right)^2
 	- \frac{c_{b2}}{\sigma} \nabla \nut \cdot \nabla \nut - \frac{1}{\sigma}\nabla \cdot \left [ (\nu + \nut) \nabla \nut \right ] = 0
\end{equation}

where $d$ is the distance to the nearest wall boundary, $\sigma = 2/3$, and $\tilde{S}$ is the modified vorticity magnitude given as,

\begin{equation*}
\tilde{S} = S + \frac{\nut}{\kappa^2d^2}f_{v2}, \qquad S = \sqrt{ 2 W(\uu) : W(\uu)},
\end{equation*}

Here $\kappa=0.4$ is the von K\'arm\'an constant, $W$ is the rotation-rate tensor, $f_{v2}$ is a damping function to enforce the logarithmic profile, with other damping functions given as:

\begin{equation*}
\label{eq:SADampingFuncs}
	\begin{split}
		f_{v1} = \frac{\chi^3}{\chi^3 + c_{v1}^3}, \qquad  \chi = \frac{\nut}{\nu}, \qquad f_{v2}= 1 - \frac{\chi}{1 + \chi f_{v1}} \qquad f_{t2} = c_{t3}e^{-c_{t4}\chi^2}\\
		f_w = g\left[ \frac{1 + c_{w3}^6}{g^6 + c_{w3}^6} \right]^{\frac{1}{6}}, \qquad g = r + c_{w2} (r^6 - r), \qquad r = \frac{\nut}{\tilde{S}\kappa^2d^2}, \\
	\end{split}
\end{equation*}

and model coefficients are specified as:

\begin{equation*}
\label{eq:SANegCoeffs}
\begin{array}{lllllllllll}
c_{b1} &= 0.1355&,\ c_{b2} &= 0.622&,\ c_{v1} &= 7.1&,\ c_{v2} &= 0.7&,\ c_{v3} &= 0.9 \\[2ex]
c_{w1} &= \dfrac{c_{b1}}{\kappa} + \dfrac{1+c_{b2}}{\sigma}&,\ c_{w2} &= 0.3&,\ c_{w3} &= 2&,\ c_{t3} &= 1.2&,\ c_{t} &= 0.5.
\end{array}
\end{equation*}

From dimensional considerations, $\nut$ is proportional to the product of characteristic length and velocity, and as a result proportional to the Reynolds number:

\begin{equation}
\nut \propto u L \sim f(Re)
\end{equation}

More details on the implementation of this model can be found in \cite{guiza2020anisotropic}, and more details on the turbulent viscosity models can be found in \cite{pope2001turbulent}. Variants of the SA model exist in the literature, most of which are collected in NASA's turbulence modeling resource webpage \cite{rumsey2010description}. In this present work, the \emph{negative Spalart-Allmaras Model} was selected due to its capability to avoid the generation of negative turbulent viscosity without the use of clipping \cite{allmaras2012modifications}. These equations were cast into a stabilized finite element formulation and solved using an in-house variational multi-scale solver CimLib CFD \cite{hachem2013immersed}. For additional details, the reader is referred to \cite{hachem2013immersed} and \cite{guiza2020anisotropic}  

\subsection{Datasets of turbulent flow around obstacle}
\label{subsec:datasets}

\begin{figure}
\centering
\def\xaxis{15}
\def\yaxis{10}
\def\sc{0.55}
\begin{subfigure}[t]{.8\textwidth}
	\centering
	\includegraphics{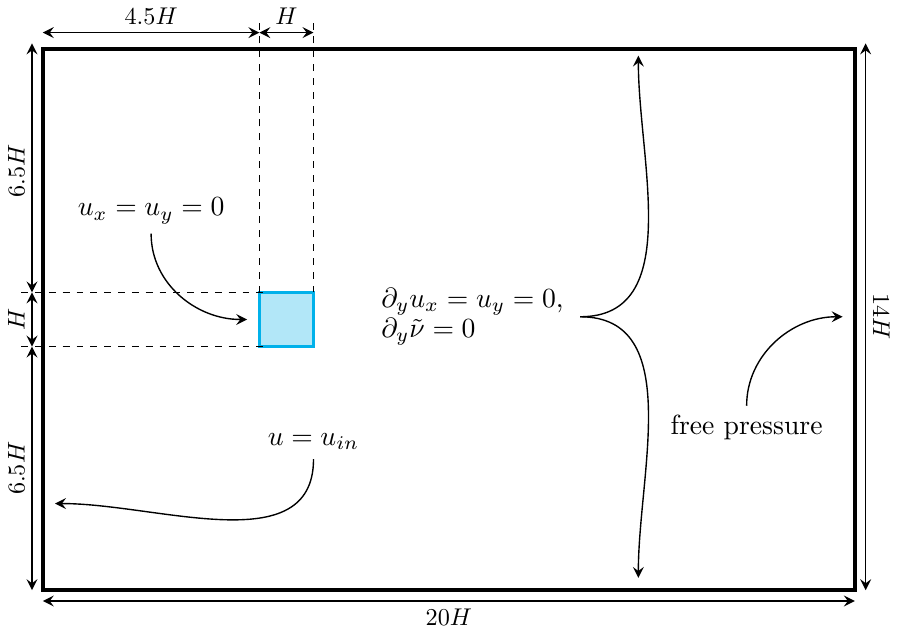}
	\caption{Sketch of the considered problem (not to scale)}
	\label{fig:cylinder_sketch}
\end{subfigure}

\medskip

\begin{subfigure}[t]{.45\textwidth}
	\centering
	\fbox{\includegraphics[height=.65\textwidth]{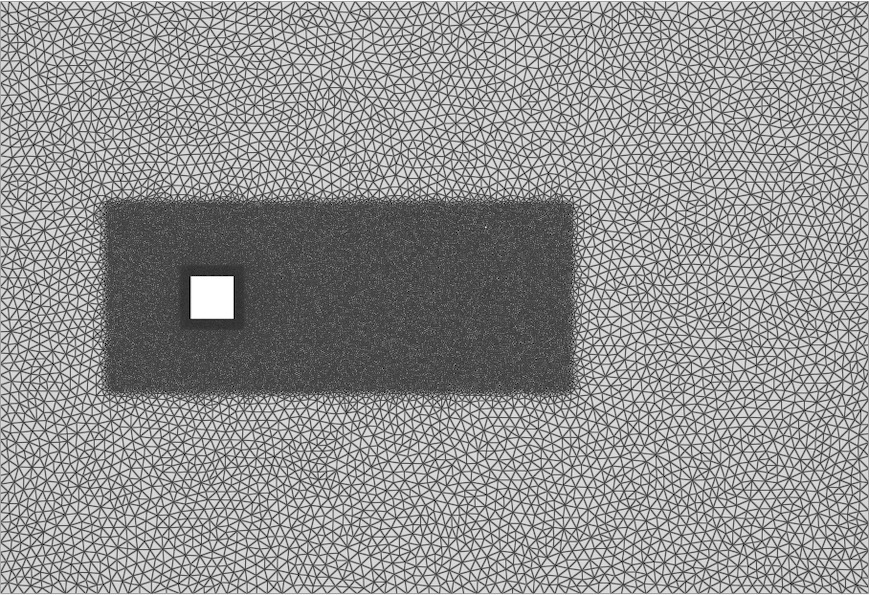}}
	\caption{Associated mesh}
	\label{fig:cylinder_mesh}
\end{subfigure} \quad
\begin{subfigure}[t]{.45\textwidth}
	\centering
	\fbox{\includegraphics[height=.65\textwidth]{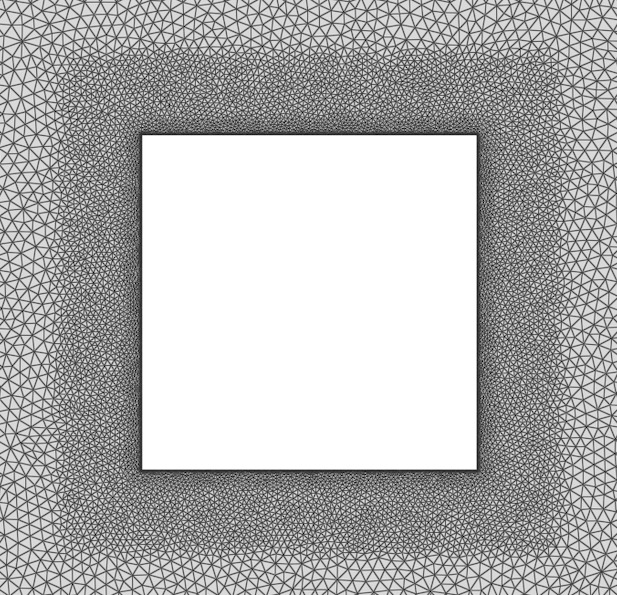}}
	\caption{Zoom on the cylinder area}
	\label{fig:cylinder_mesh_zoom}
\end{subfigure}
\caption{\textbf{2D square cylinder configuration and mesh used for the study.} (\ref{fig:cylinder_sketch}) The cylinder lateral size is denoted $H$, and is centered at the origin of the domain. The dimensions of the computational domain are $\left[ -5H, 15H \right] \times \left[ -7H, 7H \right]$ in the streamwise $x$ and crosswise $y$ directions. (\ref{fig:cylinder_mesh})-(\ref{fig:cylinder_mesh_zoom}) The mesh used for CFD computations is refined along with mesh-convergence.}
\label{fig:cylinder_setup}
\end{figure}

We consider the widely benchmarked turbulent flow past a two-dimensional (2D) square cylinder \cite{rodi1997status, guiza2020anisotropic}. A sketch of the problem, including its dimensions, is presented in figure \ref{fig:cylinder_setup}, along with the associated mesh. The baseline Reynolds number is set to $\num{22e3}$, based on the inlet velocity and the cylinder diameter. The inflow boundary conditions are $\uu=(V_{in},0)$, together with $\tilde{\nu}=3\nu$, which corresponds to a ratio of eddy to kinematic viscosity of approximately $0.2$. For the lateral boundaries, we use symmetry conditions $\partial_y u_{x}=u_{y}=0 $ and $\partial_y \tilde{\nu}=0 $. For the outflow, $\partial_x u_{x}= \partial_x u_{y}=0,$ $\partial_x \tilde{\nu}=0$ together with $p=0$ are prescribed. Finally, no-slip conditions $\uu=0 $ and $ \tilde{\nu}=0$ are imposed at the cylinder surface.  

\begin{figure}
\centering
\begin{subfigure}[b]{.3\linewidth}
	\centering
	\fbox{\includegraphics[width=.9\linewidth]{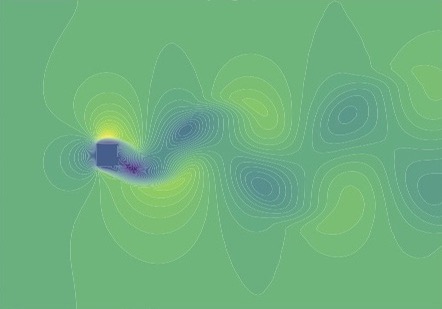}}
	\caption*{}
\end{subfigure} \quad
\begin{subfigure}[b]{.3\linewidth}
	\centering
	\fbox{\includegraphics[width=.9\linewidth]{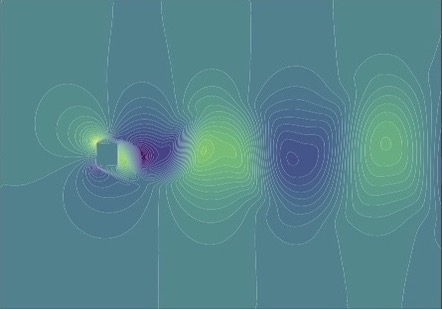}}
	\caption*{}
\end{subfigure} \quad
\begin{subfigure}[b]{.3\linewidth}
	\centering
	\fbox{\includegraphics[width=.9\linewidth]{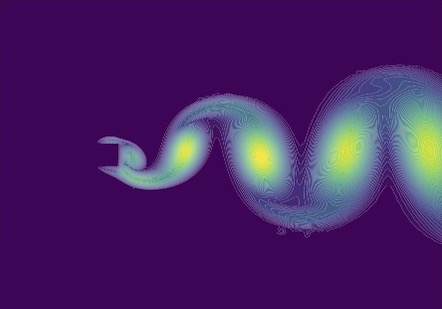}}
	\caption*{}
\end{subfigure} 

\vspace{-2em}%
\begin{subfigure}[t]{.3\linewidth}
	\centering
	\includegraphics{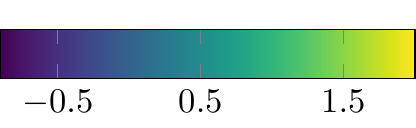}
	\caption{$u$ field}
	\label{fig:dataset_u}
\end{subfigure} \quad
\begin{subfigure}[t]{.3\linewidth}
	\centering
	\includegraphics{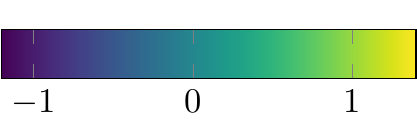}
	\caption{$v$ field}
	\label{fig:dataset_v}
\end{subfigure} \quad
\begin{subfigure}[t]{.3\linewidth}
	\centering
	\includegraphics{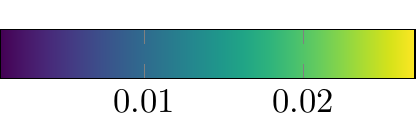}
	\caption{$\tilde{\nu}$ field}
	\label{fig:dataset_nu}
\end{subfigure}

\caption{\textbf{Snapshot of velocities $u$ (\ref{fig:dataset_u}), $v$ (\ref{fig:dataset_v}), and turbulent viscosity $\tilde{\nu}$ (\ref{fig:dataset_nu})} from dataset SqRe22k.}
\label{fig:dataset_samples}
\end{figure}

Following the problem setup and methods, a baseline dataset (hereafter referred to as SqRe22k) composed of 3000 snapshots of steady velocities and SA turbulent viscosities is generated by skipping the transient regime and storing the established regime (\red{\textit{i.e.} each snapshot is captured only after the flow is established}). Each snapshot is sampled on a rectilinear grid having spatial dimensions of $\left(N_x \times N_y\right) = \left(360 \times 300\right)$. \red{ The sampling on a rectilinear grid was performed to facilitate the use of CFD data coming from unstructured meshes. The same rectilinear grid was used to perform sampling on the square and circular obstacles. For points inside the obstacle, the velocities and turbulent viscosities were zeroed out, following the no-slip boundary conditions on the obstacle. In practice, it would be possible to skip the unstructured-to-structured sampling by making use of the graph neural networks, as presented in recent works \cite{chen2021graph}.} The dataset is deliberately not normalized to achieve robust and generalizable training. For testing purposes, additional datasets are also generated by changing the obstacle to a 2D circular cylinder, and by modifying the Reynolds number. As is summarised in table \ref{table:datasets}, six different datasets are obtained. Sample snapshots of velocity and turbulent viscosity from SqRe22k are shown in figure \ref{fig:dataset_samples}. In the following, the training subset is composed of 75\% of the SqRe22k samples and 25\% of the CyRe44k samples, while the remaining samples are reserved for the validation and testing subsets \red{(each of the latter is therefore composed of 12.5\% of the SqRe22k and 37.5\% of the CyRe44k)}.

\begin{table}
\footnotesize
\caption{\textbf{Datasets generated for the present study.} Two types of obstacles and three different Reynolds numbers are considered, resulting in six different datasets, each holding 3000 snapshots of steady-state velocities and turbulent velocities.}
\label{table:datasets}
\centering
\medskip
\begin{tabular}{ccc}
\toprule
\textbf{Dataset name}	& \textbf{Re}	& \textbf{Obstacle type} 	\\\midrule
SqRe22k				& \num{22e3}	& 2D square			\\\midrule
SqRe44k				& \num{44e3}	& 2D square			\\\midrule
SqRe88k				& \num{88e3}	& 2D square			\\\midrule
CyRe22k				& \num{22e3}	& 2D cylinder			\\\midrule
CyRe44k				& \num{44e3}	& 2D cylinder			\\\midrule
CyRe88k				& \num{88e3}	& 2D cylinder			\\\bottomrule
\end{tabular}
\bigskip
\end{table}

\section{Network architecture and training procedure}

\subsection{Deep learning model}

Given the input dataset $\bm{x}$ (here, the velocity snapshots from the RANS simulations) and the desired output dataset $\bm{y}$ (here, the turbulent viscosity snapshots from the RANS simulations), we desire to find the optimal set of weights and biases $\bm{\theta} = \left(\bm{w}, \bm{b}\right)$ in a deep-learned model $f$ such that $f(\bm{x};\bm{\theta}) = \bm{y}$. The set of free parameters $\bm{\theta}$ is optimized using Adam \cite{kingma2014adam}, in order to iteratively minimize the mean squared error (MSE) loss defined as: 

\begin{equation}
\mathcal{L} = \frac{1}{n_s} \sum_{i=1}^{n_s} (\bm{y}^{i}  - f(\bm{x};\bm{\theta})^{i})^2,
\label{lossRelation}
\end{equation}

where $n_s$ is the number of samples. The full training dataset is shown repeatedly to the network after a shuffling step during the training, and each pass is referred to as an \textit{epoch}. An early stopping criterion is used along with a reduction of learning rate if learning doesn't improve after every 100 epochs. The neural network was implemented using TensorFlow \cite{abadi2016tensorflow}, and trained on an Nvidia Tesla V100 GPU.

\begin{figure}
\centering
\includegraphics{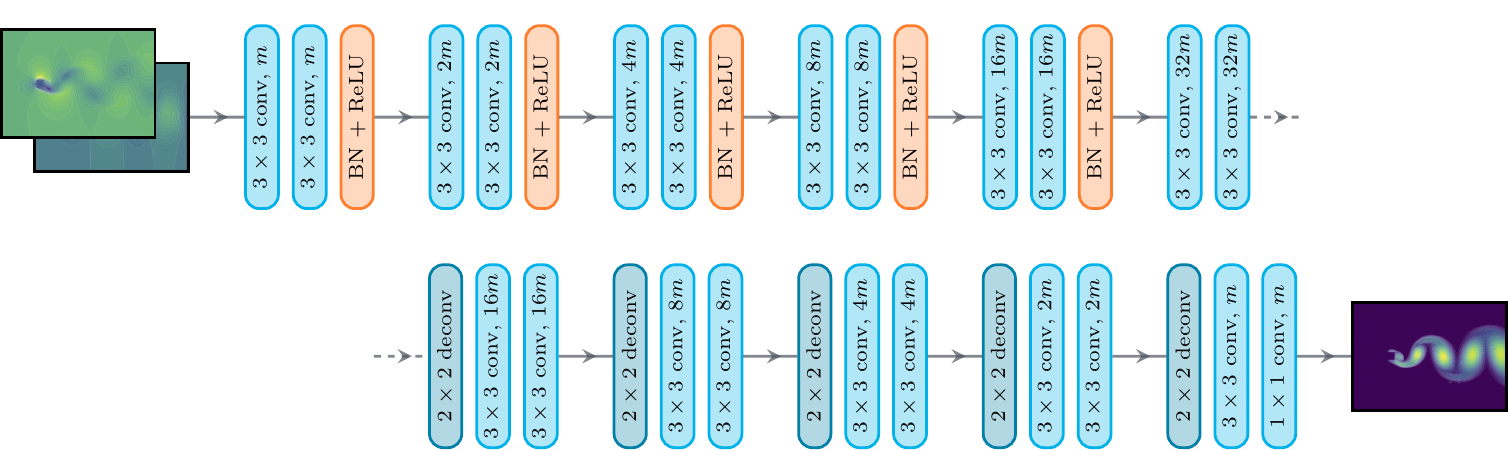}

\caption{\textbf{Proposed auto-encoder network architecture.} The encoder branch is based on a convolution-convolution-batch-normalization pattern: the first convolution has a stride of $s=1$, while the second has a stride of $s=2$. The batch-normalization layer is followed by a rectified linear unit (ReLU) layer. At each occurrence of the pattern, the spatial dimensions are divided by two, while the number of filters, noted $m$, is doubled. In the decoder branch, a transposed convolution step is first applied to the input from the previous layer, while the number of filters is halved and two convolution layers are applied. At the end of the last layer, a $1 \times 1$ convolution is applied to obtain the final output.}
\label{fig:autoencoder}
\end{figure}

The network architecture proposed for the present work is an \textit{auto-encoder} structure \cite{hinton2006reducing}. Auto-encoders contain two parts: (i) a converging part that decreases the spatial dimension of the input (the encoder) and compresses the input using successive convolutions, and (ii) a diverging part that rebuilds a predicted output of the same size as input (the decoder). The encoder and decoder handle the spatial-dimensionality reduction by compressing the high-dimensional spatial data, using convolutional layers, to a low-dimensional representation called latent space. For example, a $N_{y} \times N_{z}$ feature map can be reduced to $N_y$/2 $\times$ $N_z$/2 using a convolutional layer with a stride of 2. An essential aspect of this operation is that it preserves the most important features of the map. To increase robustness and generalization of the trained model, data standardization was not performed. Instead, batch normalization layers were used, which apply a transformation that maintains the mean and standard deviation of output close to 0 and 1, respectively. The proposed network architecture is shown in figure \ref{fig:autoencoder}. In the literature, similar architectures (trained with full-scale inputs) were successfully exploited for studies focusing on turbulent flows \cite{fukami2019synthetic, mohan2019compressed}.

The convolutional filters used in the proposed architecture incorporate a symmetric boundary condition into the padding operation. Classically, padding is used to preserve the spatial dimensions of the field being convoluted, but the standard zero-padding approach doesn't usually represent the expected physical behavior. \red{Indeed, padding with zeros everywhere would violate the representation of existing boundary conditions, for example, the notion of wall-boundaries would have lesser significance if a region is padded with zeros on all the sides in a channel flow \cite {patil2019development} }.  To preserve the boundary conditions after multiple successive convolutions, a boundary condition formulation was implemented such that the walls could be padded with zeros if required, while the periodic sides could be padded with adequate values from the periodic cells. The ReLU function was used as an activation function, which is known to be an effective tool for stabilizing the weight update in the machine learning process \cite{nair2010rectified}.

\subsection{Patch-based training procedure}
\label{section:patchedtrain}

\begin{figure}
\centering
\begin{subfigure}[b]{.45\linewidth}
	\centering
	\fbox{\includegraphics[width=.9\linewidth]{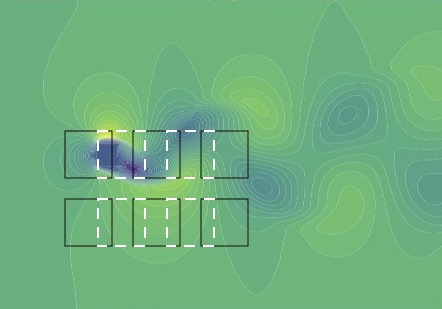}}
	\caption{Original snapshot}
	\label{fig:original_snapshot}
\end{subfigure} \quad
\begin{subfigure}[b]{.45\linewidth}
	\centering
	\begin{subfigure}[b]{.3\linewidth}
		\centering
		\fbox{\includegraphics[width=.9\linewidth]{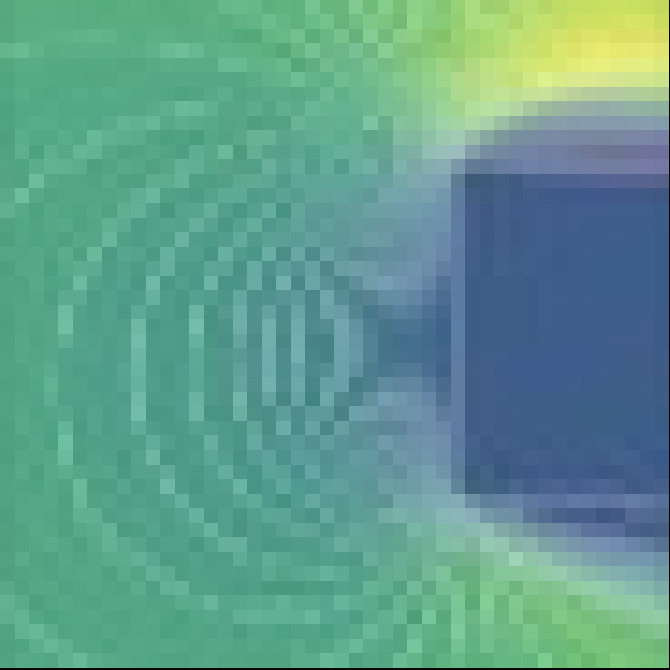}}
		\caption*{}
	\end{subfigure}
	\begin{subfigure}[b]{.3\linewidth}
		\centering
		\fbox{\includegraphics[width=.9\linewidth]{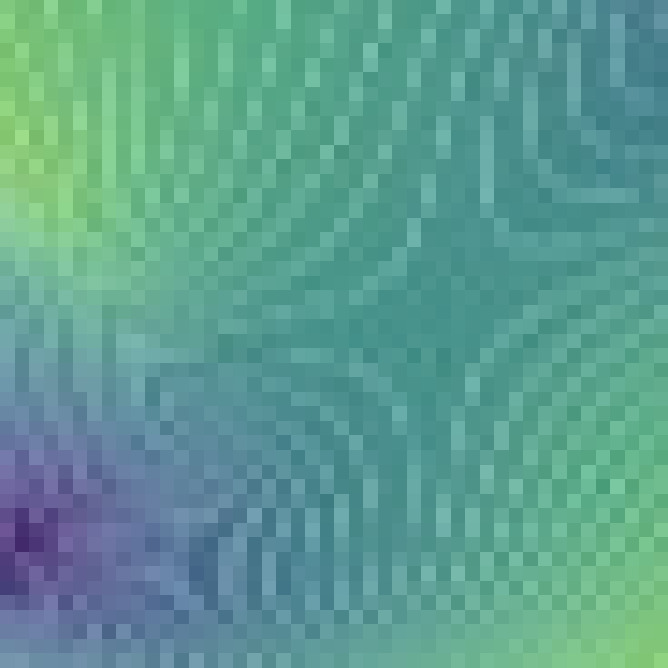}}
		\caption*{}
	\end{subfigure}
	\begin{subfigure}[b]{.3\linewidth}
		\centering
		\fbox{\includegraphics[width=.9\linewidth]{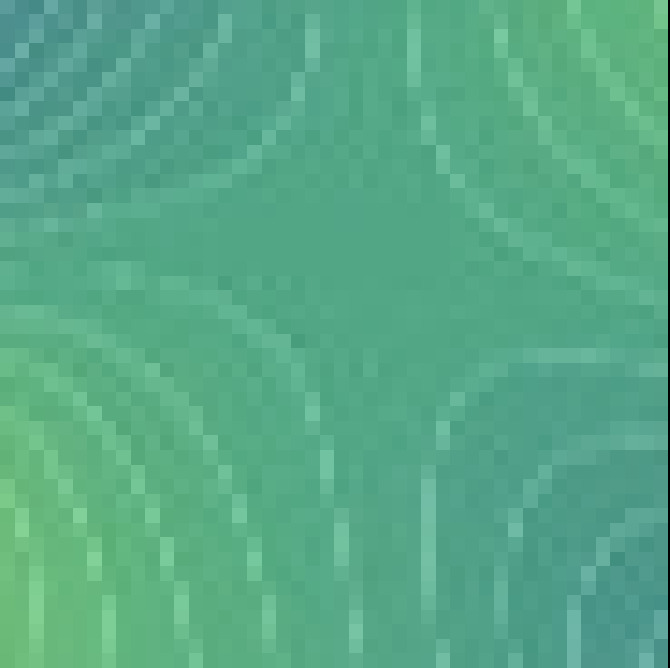}}
		\caption*{}
	\end{subfigure}

	\vspace{-1em}
	\begin{subfigure}[b]{.3\linewidth}
		\centering
		\fbox{\includegraphics[width=.9\linewidth]{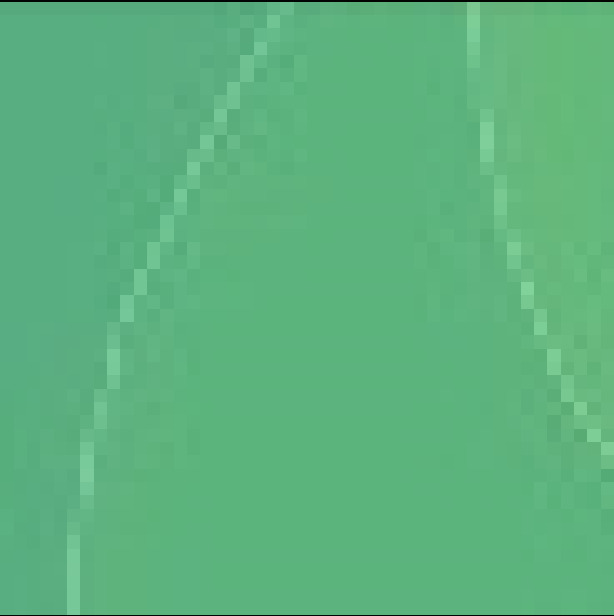}}
		\caption*{}
	\end{subfigure}
	\begin{subfigure}[b]{.3\linewidth}
		\centering
		\fbox{\includegraphics[width=.9\linewidth]{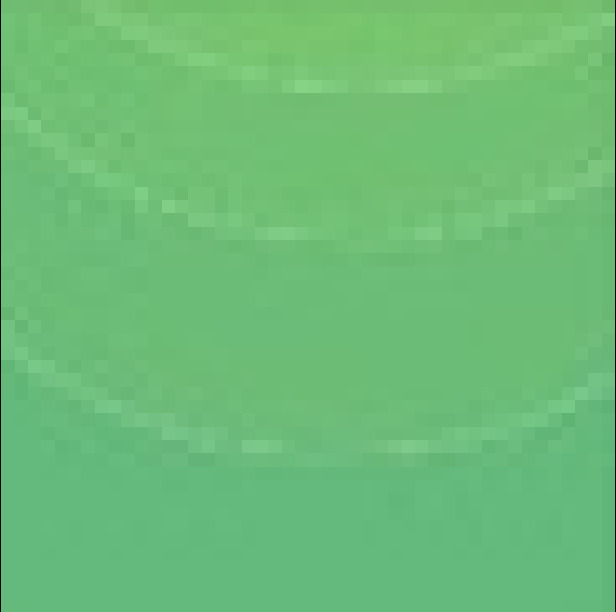}}
		\caption*{}
	\end{subfigure}
	\begin{subfigure}[b]{.3\linewidth}
		\centering
		\fbox{\includegraphics[width=.9\linewidth]{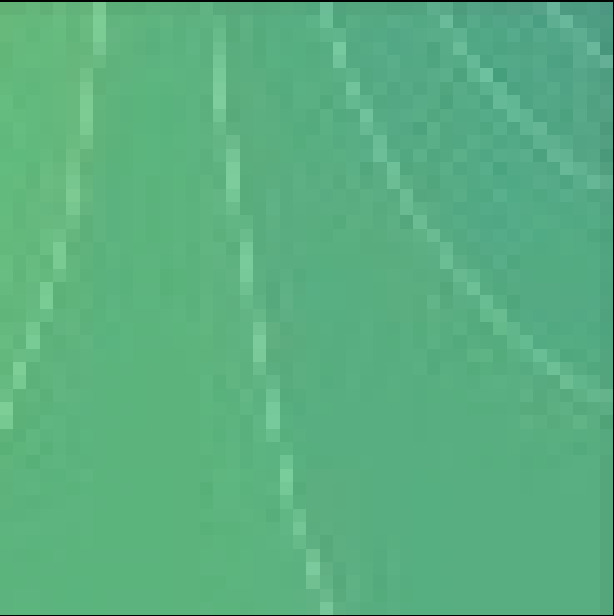}}
		\caption*{}
	\end{subfigure}%
	\vspace{-1.1em}%
	\caption{Extracted patches}
	\label{fig:patches}
\end{subfigure}
\caption{\textbf{Patch extraction from $u$ field}. Patches in figure (\ref{fig:patches}) are obtained from the original snapshot (\ref{fig:original_snapshot}) \red{For better clarity of the figure, overlapping is only applied in the horizontal direction, and different colors are used to differentiate overlapping patches.} Similarly, patches at the same corresponding locations are taken for $v$ and $\tilde{\nu}$ fields.}
\label{fig:patch_extraction}
\end{figure}

We remind the goal of the present work, which is to train a deep learning model to infer the turbulent viscosity $\tilde{\nu}$ at every grid point from the velocities $(u,v)$ at the same position. As underlined earlier, no data-preprocessing tasks such as normalization or standardization were used, and the input-output fields were used "as is" from the RANS simulation output. Similar to splitting between training and validation dataset as described in section \ref{subsec:datasets}, we use a mixture of the SqRe22k and CyRe44k datasets. The first stage of patch-based learning consists of dividing each snapshot of the dataset into smaller $n \times n$ overlapping patches with stride $s$, as is shown in figure \ref{fig:patch_extraction}. In this case, the number of patches obtained can be doubled by considering an up-down flipping transformation on the same snapshot.

\begin{figure}
\centering
\includegraphics{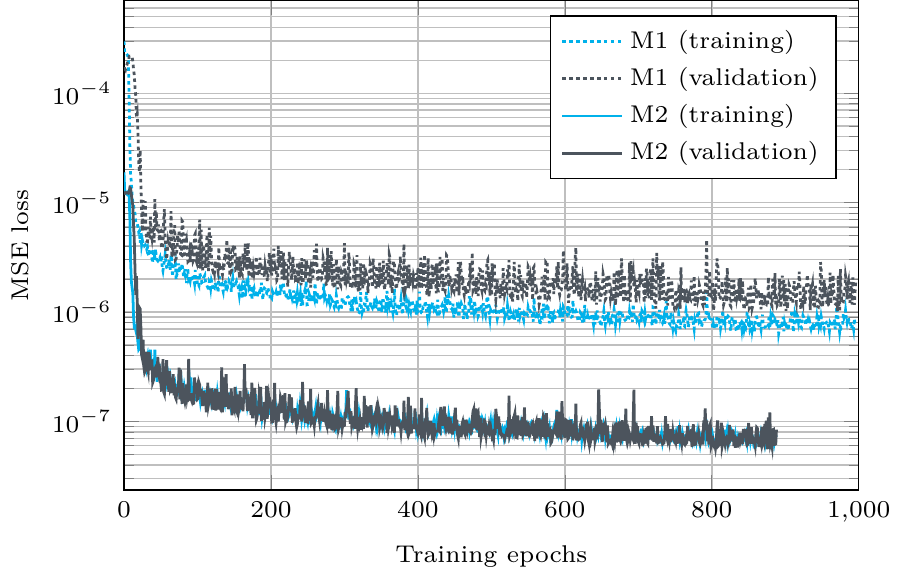}
%
%
%
\caption{\textbf{Training and validation loss history} for the M1 and M2 training methods. The patch-based technique (M2) yields lower error and better generalization than the baseline M1, as evidenced by the negligible gap between validation and training curves. M1 training was performed for 1000 epochs, and the M2 training was stopped after 850 epochs when error stopped improving.}
\label{fig:training_history}
\end{figure}

For a baseline comparison, the proposed network is also trained conventionally over the full-spatial field dimensions, without using patch-based learning (this training method is hereafter referred to as M1). In this context, the batch size is 32, and the learning rate is $0.001$. When a sufficient accuracy level is reached and no more improvement is observed, the training is terminated using the early-stopping criterion. A decent accuracy after convergence is obtained for both training and validation subsets, with a mean-squared error of \num{1e-6}, as presented on the learning curve in figure \ref{fig:training_history}. Total training time is 0.85 hours on a Tesla V100 GPU card, for 28 million degrees of freedom.

For patch-based training, patches from the different samples are randomly shuffled together and presented to the network in batches of size 32, with a learning rate equal to 0.001 (this training method is hereafter referred to as M2). Baseline values for the patch size $n$ and the stride $s$ are chosen to be 50 and 75, respectively, but their respective impact on the training performance is evaluated in section \ref{section:results}. Similarly, the impact of batch size is assessed in the following section. The model is trained for 850 epochs, after which the accuracy stops improving, resulting in a final MSE error of the order of \num{1e-7}, \ie one order of magnitude lower than that of method M1. Total training time is 2.38 hours for 1.7 million degrees of freedom. Although this represents about 3 times the training time of method M1, it must be noticed that the final M2 accuracy is significantly lower than that of M1, as is visible in figure \ref{fig:training_history}. More, the final generalization level is also superior, evidenced by the negligible gap between validation and training curves. As the patch-based approach grounds the learning in a local velocity-to-turbulent-viscosity inference, it is argued that the trained network is able to re-use local mappings from one snapshot to another, leading to improved generalization capabilities compared to a monolithic snapshot-to-snapshot inference.

\section{Results and discussion}
\label{section:results}

\begin{figure}
\centering
\shifttext{-14mm}{\includegraphics{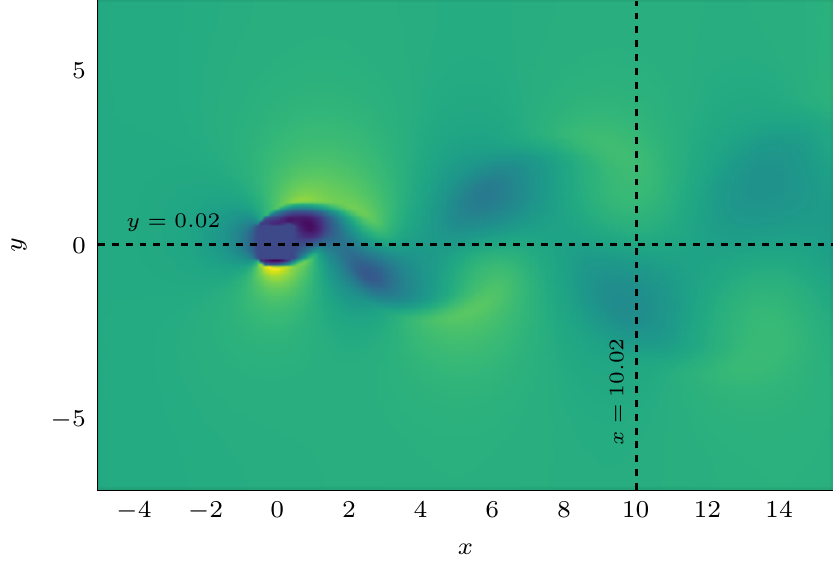}}
%
\caption{\textbf{Locations of the probe lines used for comparison to CFD reference}.}
\label{fig:line_locations} 
\end{figure}

In this section, the benefits induced by the patch-based training procedure are compared with that of the regular M1 training method on predictive tasks. To this end, predictions of both models are evaluated against reference solutions obtained from the CFD solver. In the remaining of this section, training data consists of 75\% of samples from the SqRe22k dataset and 25\% of samples from the CyRe44k dataset. Such a mixing of datasets is used to assess the generalization capabilities of the two methods, as both datasets present similar flow features, but with different obstacles. First, comparisons are made on out-of-training samples from the SqRe22k dataset using baseline training parameters. Then, predictions obtained with snapshots from different datasets (SqRe44k, SqRe88k, CyRe22k, and CyRe88k) are evaluated against their references. Finally, a parametric study considering the impact of batch size $b$, the patch size $n$, and the stride size $s$ on the final performance is proposed. Overall, comparisons are made on the basis of (i) contour plots of predicted and expected $\tilde{\nu}$, (ii) 1D plots of $\tilde{\nu}$ along streamwise and spanwise lines at different locations in the domain, as shown in figure \ref{fig:line_locations}, and (iii) scatter and density plots of the predicted $\tilde{\nu}$ against reference values.

\subsection{Comparison on out-of-training snapshot}

\begin{figure}
\centering
\begin{subfigure}[b]{.4\linewidth}
	\centering
	\shifttext{-12mm}{\includegraphics{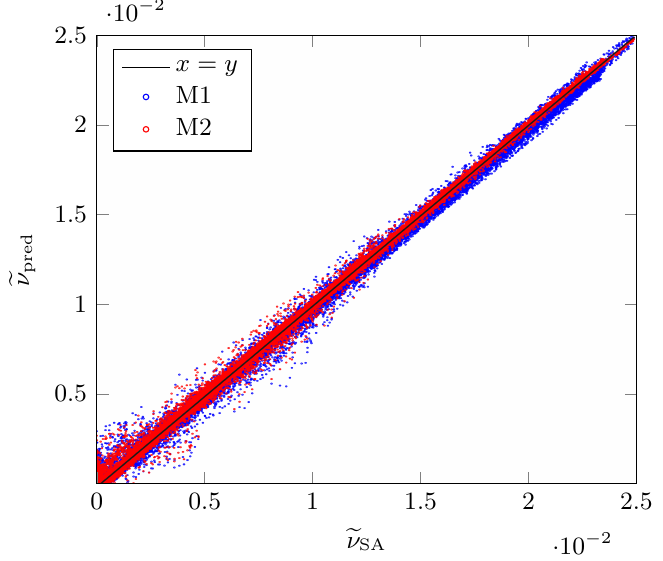}}
%
	\caption{ }
	\label{fig:scatter}
\end{subfigure} \qquad \quad
\begin{subfigure}[b]{.4\linewidth}
	\centering
	\shifttext{-16mm}{\includegraphics{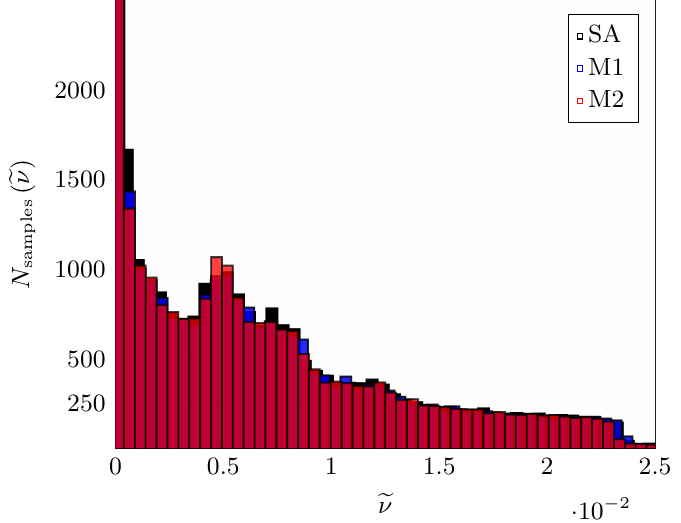}}
	\caption{}
	\label{fig:hist}
\end{subfigure}
\caption{\textbf{Scatter plot and histogram} of predicted and expected $\tilde{\nu}$ for an out-of-training snapshot of SqRe22k. (\ref{fig:scatter}) The plot is a superposition of two scatter plots, namely SA against M1 and SA against M2. \red{(\ref{fig:hist}): The histogram compares the occurence of truth and predictions on a step-type filled histogram.}}
\label{fig:scatter_hist}
\end{figure}

\begin{figure}
\centering
\begin{subfigure}[b]{.45\linewidth}
	\centering
	\fbox{\includegraphics[width=.9\linewidth]{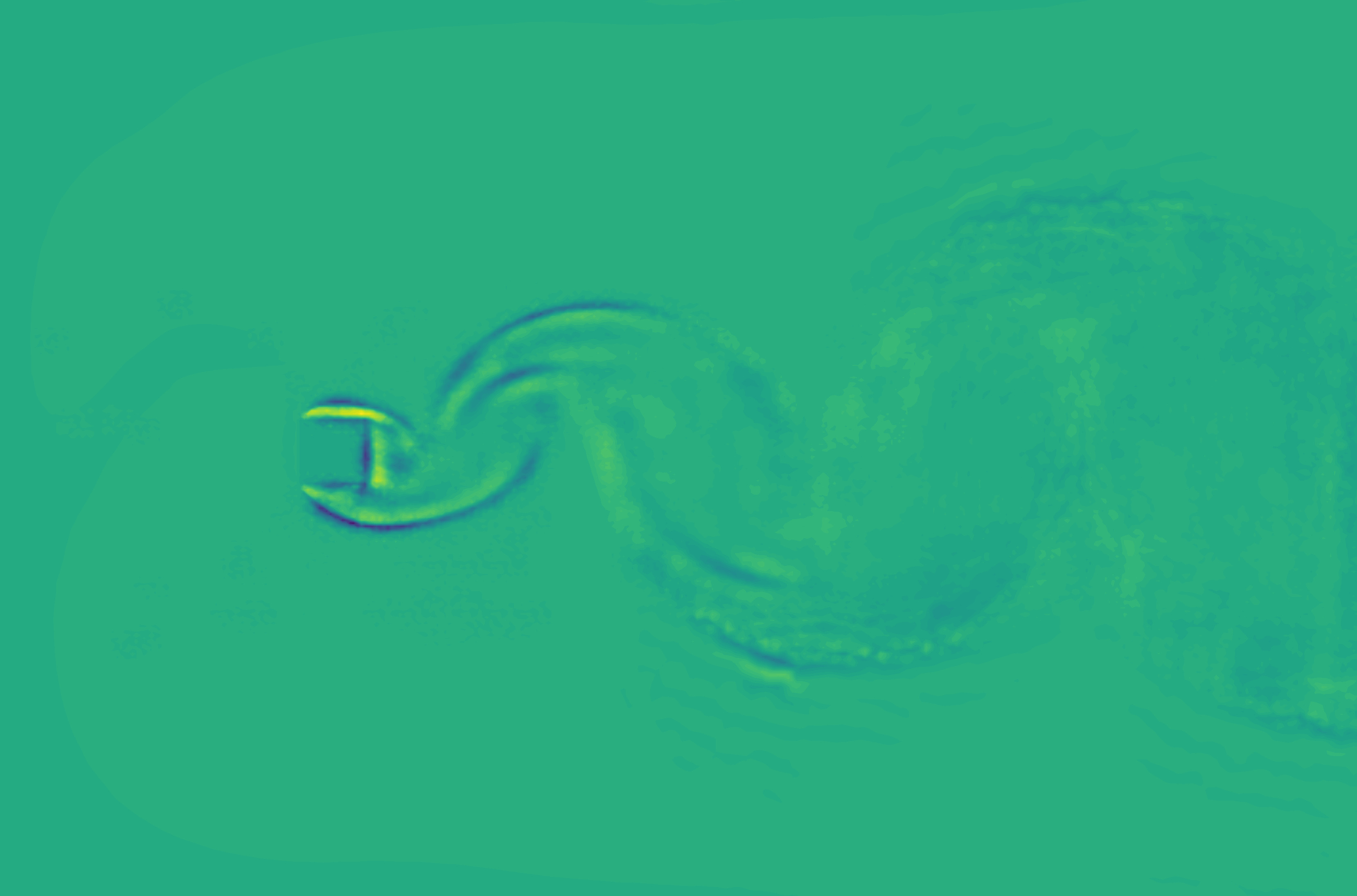}}
	\caption*{}
\end{subfigure} \quad
\begin{subfigure}[b]{.45\linewidth}
	\centering
	\fbox{\includegraphics[width=.9\linewidth]{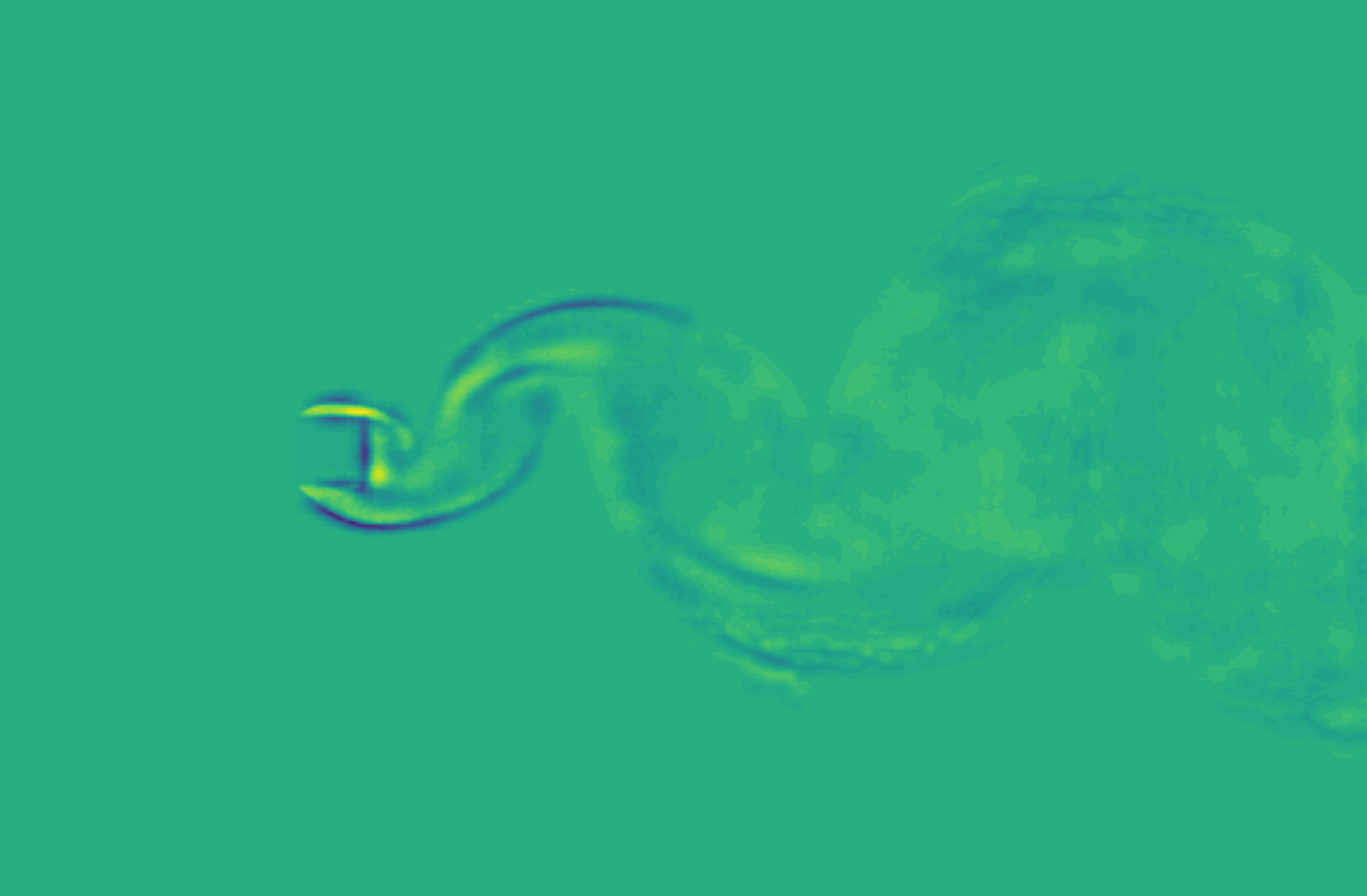}}
	\caption*{}
\end{subfigure}

\vspace{-2em}%
\hspace{0.2em}%
\begin{subfigure}[t]{.45\linewidth}
	\centering
	\includegraphics{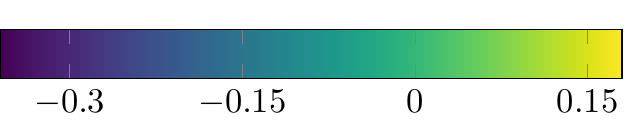}
	\caption{Relative error for prediction from M1}
	\label{fig:error_left}
\end{subfigure} \quad
\begin{subfigure}[t]{.45\linewidth}
	\centering
	\includegraphics{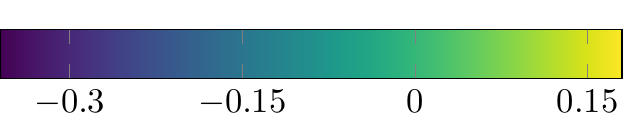}
	\caption{Relative error for prediction from M2}
	\label{fig:error_right}
\end{subfigure}
\caption{\textbf{Contour plots of relative errors} obtained from the same snapshot input from dataset SqRe22k, using methods M1 (\ref{fig:error_left}) and M2 (\ref{fig:error_right}). In both cases, maximal error levels are observed in the vicinity of the obstacle.}
\label{fig:error_fields}
\end{figure}

In this section and the following, baseline training parameters are used, \ie batch size is equal to 32, patch size $n$ is equal to 50, and stride size $s$ is equal to 75. As stated above, the training data consists of 75\% of samples from the SqRe22k dataset and 25\% of samples from the CyRe44k dataset. M1 and M2 models' predictive capabilities are compared on an out-of-training snapshot from the SqRe22k dataset, as shown in figure \ref{fig:scatter_hist}. As can be observed on the scatter plot (figure \ref{fig:scatter}), both M1 and M2 methods are in good accordance with the reference regarding the predicted $\tilde{\nu}$. Still, the M2 prediction presents an average relative deviation of 2.25\% on the entire sample, against 5.04\% for M1. More, its maximum relative deviation is also lower, with 36.44\% for M2, against 76.23\% for M1. To illustrate, the error fields obtained with M1 and M2 predictions are shown on the same snapshot in figure \ref{fig:error_fields}. 

\subsection{Comparison on out-of training datasets}

\begin{figure}
\centering
\begin{subfigure}[b]{\linewidth}
\centering
\begin{subfigure}[b]{.4\linewidth}
	\centering
	\shifttext{-10mm}{\includegraphics{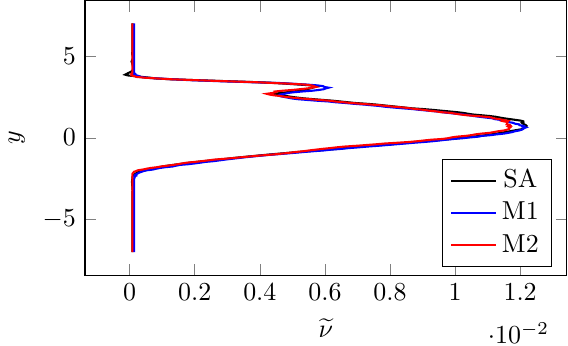}}
	\caption{SqRe44k}
	\label{fig:pred_x_SqRe44k}
\end{subfigure} \quad
\begin{subfigure}[b]{.4\linewidth}
	\centering
	\shifttext{-10mm}{\includegraphics{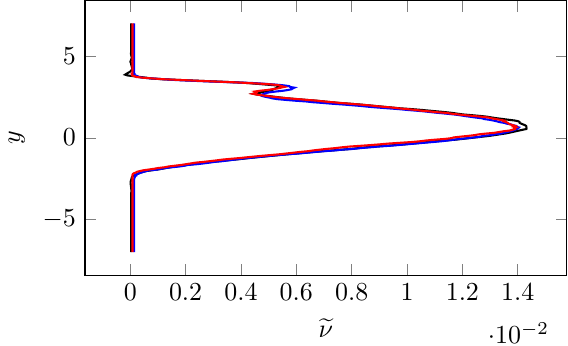}}
	\caption{SqRe88k}
	\label{fig:pred_x_SqRe88k}
\end{subfigure}

\medskip

\begin{subfigure}[b]{.4\linewidth}
	\centering
	\shifttext{-10mm}{\includegraphics{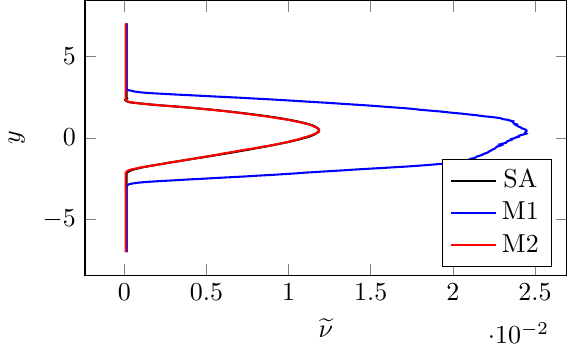}}
	\caption{CyRe22k}
	\label{fig:pred_x_CyRe22k}
\end{subfigure} \quad
\begin{subfigure}[b]{.4\linewidth}
	\centering
	\shifttext{-8mm}{\includegraphics{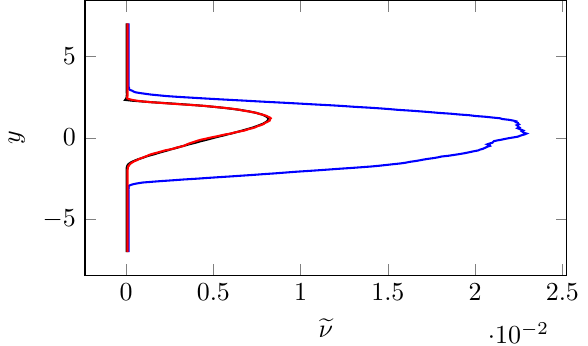}}
	\caption{CyRe88k}
	\label{fig:pred_x_CyRe88k}
\end{subfigure}

\medskip

\caption*{(a)-(d): $\widetilde{\nu}$ in the span-wise direction at $x = 10.02$}
\label{fig:x_cst}
\end{subfigure}

\medskip
\medskip
\medskip

\begin{subfigure}[b]{\linewidth}
\centering
\begin{subfigure}[b]{.4\linewidth}
	\centering
	\shifttext{-8mm}{\includegraphics{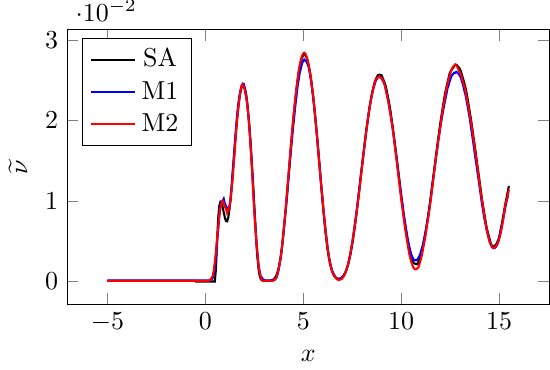}}
	\caption{SqRe44k}
	\label{fig:pred_y_SqRe44k}
\end{subfigure} \quad
\begin{subfigure}[b]{.4\linewidth}
	\centering
	\shifttext{-8mm}{\includegraphics{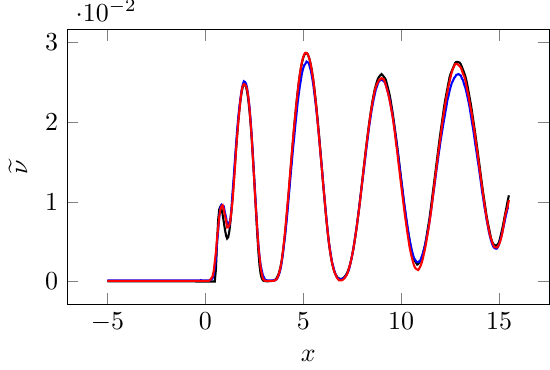}}
	\caption{SqRe88k}
	\label{fig:pred_y_SqRe88k}
\end{subfigure}

\medskip

\begin{subfigure}[b]{.4\linewidth}
	\centering
	\shifttext{-8mm}{\includegraphics{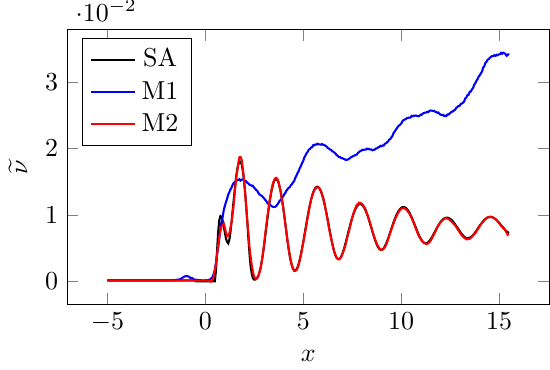}}
	\caption{CyRe22k}
	\label{fig:pred_y_CyRe22k}
\end{subfigure} \quad
\begin{subfigure}[b]{.4\linewidth}
	\centering
	\shifttext{-8mm}{\includegraphics{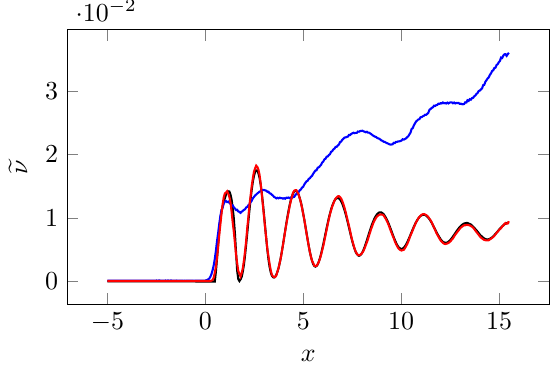}}
	\caption{CyRe88k}
	\label{fig:pred_y_CyRe88k}
\end{subfigure}

\medskip

\caption*{(e)-(h): $\widetilde{\nu}$ in the stream-wise direction at $y = 0.02$}
\label{fig:y_cst}
\end{subfigure}

\caption{\textbf{Line plots along $x = 10.2$ and along $y = 0.1$} comparing prediction accuracies of M1 and M2 on out-of-training samples from datasets SqRe44k (\ref{fig:pred_x_SqRe44k})-(\ref{fig:pred_y_SqRe44k}), SqRe88k (\ref{fig:pred_x_SqRe88k})-(\ref{fig:pred_y_SqRe88k}), CyRe22k (\ref{fig:pred_x_CyRe22k})-(\ref{fig:pred_y_CyRe22k}) and CyRe88k (\ref{fig:pred_x_CyRe88k})-(\ref{fig:pred_y_CyRe88k}). M1 and M2 perform similarly on datasets with a square obstacle, even on higher $Re$ values. Yet, M1 consistently fails at predicting accurate $\tilde{\nu}$ on samples with cylindrical obstacle, while M2 presents an almost-perfect fit with CFD reference. \red{The small deviation observed for M2 at the top of the square cylinder can be likely attributed to the unstructured-to-structured data sampling, and its study is deferred to a future work.}}
\label{fig:axis_nu}
\end{figure}

\begin{figure}
\centering
\begin{subfigure}[b]{.3\linewidth}
	\centering
	\fbox{\includegraphics[width=.9\linewidth]{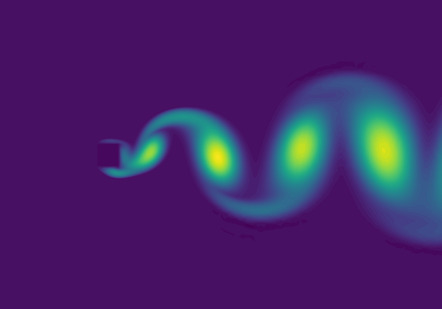}}
	\caption*{}
\end{subfigure} \quad
\begin{subfigure}[b]{.3\linewidth}
	\centering
	\fbox{\includegraphics[width=.9\linewidth]{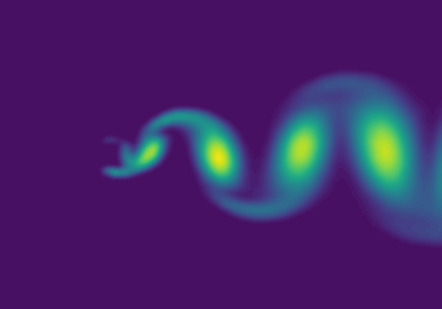}}
	\caption*{}
\end{subfigure} \quad
\begin{subfigure}[b]{.3\linewidth}
	\centering
	\fbox{\includegraphics[width=.9\linewidth]{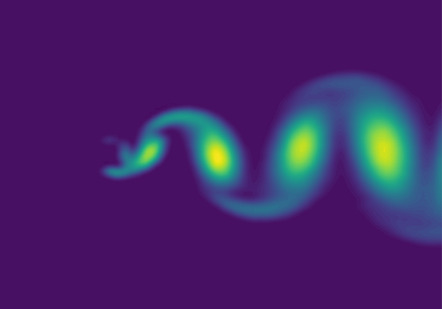}}
	\caption*{}
\end{subfigure}

\begin{subfigure}[b]{.3\linewidth}
	\centering
	\fbox{\includegraphics[width=.9\linewidth]{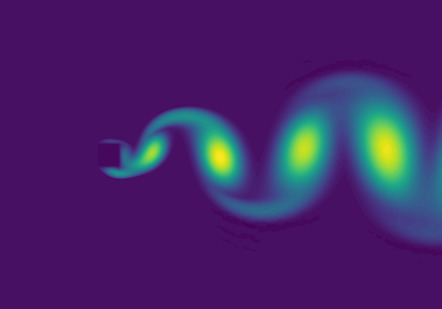}}
	\caption*{}
\end{subfigure} \quad
\begin{subfigure}[b]{.3\linewidth}
	\centering
	\fbox{\includegraphics[width=.9\linewidth]{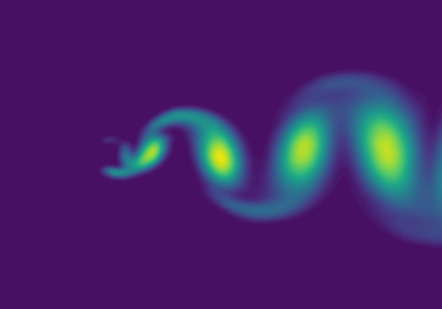}}
	\caption*{}
\end{subfigure} \quad
\begin{subfigure}[b]{.3\linewidth}
	\centering
	\fbox{\includegraphics[width=.9\linewidth]{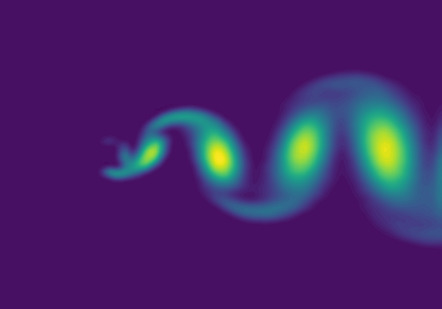}}
	\caption*{}
\end{subfigure}

\begin{subfigure}[b]{.3\linewidth}
	\centering
	\fbox{\includegraphics[width=.9\linewidth]{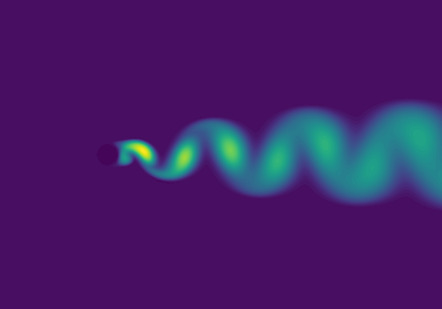}}
	\caption*{}
\end{subfigure} \quad
\begin{subfigure}[b]{.3\linewidth}
	\centering
	\fbox{\includegraphics[width=.9\linewidth]{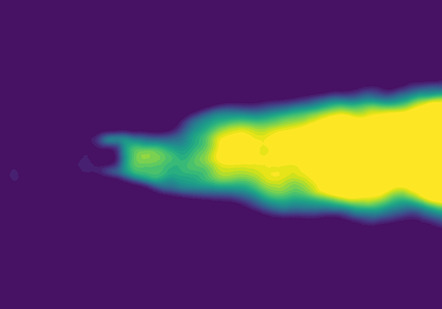}}
	\caption*{}
\end{subfigure} \quad
\begin{subfigure}[b]{.3\linewidth}
	\centering
	\fbox{\includegraphics[width=.9\linewidth]{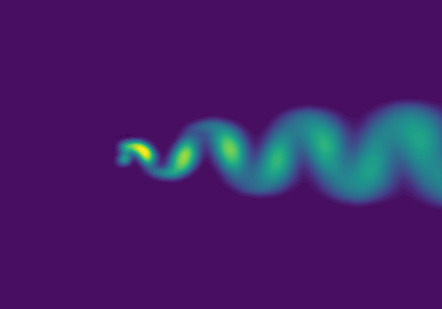}}
	\caption*{}
\end{subfigure}

\begin{subfigure}[b]{.3\linewidth}
	\centering
	\fbox{\includegraphics[width=.9\linewidth]{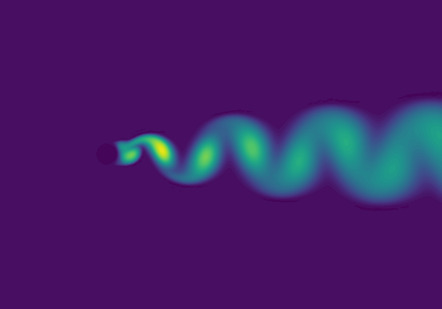}}
	\caption*{SA}
\end{subfigure} \quad
\begin{subfigure}[b]{.3\linewidth}
	\centering
	\fbox{\includegraphics[width=.9\linewidth]{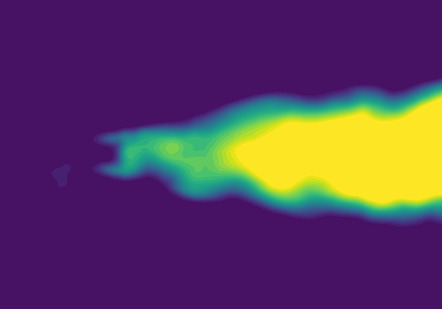}}
	\caption*{M1}
\end{subfigure} \quad
\begin{subfigure}[b]{.3\linewidth}
	\centering
	\fbox{\includegraphics[width=.9\linewidth]{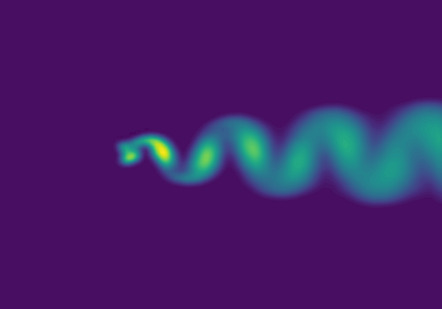}}
	\caption*{M2}
\end{subfigure}

\bigskip
\bigskip

\vspace{-2em}%
\begin{subfigure}[t]{.9\linewidth}
	\centering
	\includegraphics{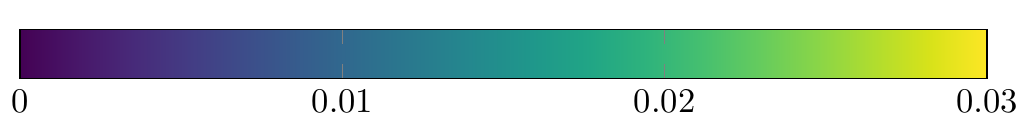}
	\caption*{}
\end{subfigure}

\caption{\textbf{Comparison of M1 and M2 $\tilde{\nu}$ predictions against CFD reference on snapshots from different out-of-training datasets}, namely SqRe44k (top row), SqRe88k (second row), CyRe22k (third row), and CyRe88k (bottom row). While M1 and M2 perform similarly on snapshots with square obstacle even at high $Re$ numbers, M1 predictions on cylindrical obstacle are significantly saturated in the wake region, showing that the model was unable to learn features from the related samples in the training dataset. Conversely, M2 predictions are in line with the SA reference.}
\label{fig:cfd_comparison}
\end{figure}

In this section, models M1 and M2 (trained on a mixed dataset composed of samples from SqRe22k and CyRe44k) are used to make predictions on snapshots from datasets SqRe44k, SqRe88k, CyRe22k, and CyRe88k, which were not used for training. M1 and M2 predictions for one snapshot of each dataset are compared against CFD reference on stream-wise and span-wise 1D plots of $\tilde{\nu}$, at the locations presented in figure \ref{fig:line_locations}. Results are shown in figure \ref{fig:axis_nu}. As can be observed, the patch-based trained model consistently outperforms the M1 model, while presenting an excellent agreement with reference data. On the $x=10.02$ line, \red{which represents full developed wake region}, performances of M1 and M2 models are close on SqRe44k and SqRe88k datasets, but M1 significantly overestimates the $\tilde{\nu}$ values on the CyRe22k and CyRe88k datasets, indicating that model M1 is unable to fully leverage the diversity of the training dataset, and only learns full-scale velocity-turbulent viscosity patterns. Conversely, the M2 model here proves its ability to learn local feature mapping from velocity field to turbulent viscosity field and accurately reconstructs it, independently of the obstacle-type and Reynolds number. Similarly, on the $y=0.02$ line, \red{which passes through the obstacle boundaries as well as the wake regions}, M1 and M2 models show similar performances on datasets with a square obstacle, while M1 largely deviates from the reference data on snapshots coming from datasets with a cylindrical obstacle. Contrarily, the M2 model again provides accurate predictions. The latter results are further emphasized on the contour plots of figure \ref{fig:cfd_comparison}, where M1 predictions on cylindrical obstacles present inaccurate features and saturated fields in the turbulent area downstream of the obstacle. This again indicates the inability of training procedures on full-scale samples to infer proper mapping from velocity fields to turbulent viscosity fields at the local scale, which is not the case of patch-based training. \red{The Reynolds numbers are of similar orders in magnitudes which explains the capabilities of M1 and M2 to extrapolate on $Re$ values outside of their training datasets.} Hence, the extrapolation capabilities of the M2 model could be assessed even at higher Reynolds number.

\def\sc{0.75}

\subsection{Parametric study}

A parametric study is performed to explore the impact of the batch size $b$, the patch size $n$ and the stride size $s$ on the MSE error $\mathcal{L}_{MSE}$  (as defined in equation (\ref{lossRelation})) computed on validation data. To this end, the performances of various $(n,s)$ pairs with relation $s = 1.5 \times n$ are first compared in terms of final validation performance and training time. To select the best performance of each pair, early stopping is used during training, and the average validation error over the last 50 epochs, noted $\overline{\mathcal{L}_{MSE}}$, is retained. As shown in table \ref{mse_ns}, the pairs $(100,150)$ and $(50,75)$ yield close performances in terms of final MSE error. Although the $(100,150)$ is slightly better in accuracy and training time, the $(50,75)$ pair is preferred for its larger amount of patches per snapshot. The larger errors of the $(20,30)$, $(10,15)$, $(6,9)$, and $(2,3)$ pairs can be attributed to the low number of points per patch making it difficult to train the model with the same hyper-parameters, while the $(200,300)$ pair prevents the efficient learning of local features, and is likely to present the same flaws as method M1.

In a second time, we consider the impact of varying stride size $s$ for the previously-select $n$ value, equal to 50. Results are presented in table \ref{mse_s}. As can be seen, no significant difference is observed for stride values ranging from 30 to 300, indicating that in this context, the amount of patches per snapshot (and thereby total samples) is not a limitation. Finally, the effect of varying batch size is assessed for $(n,s) = (50,75)$. As shown in table \ref{mse_b}, small batch sizes 8, 16, and 32 yield close error levels, while larger batch sizes are associated with errors larger by roughly one order of magnitude. Although $b=8$ is slightly lower than the other values, $b=32$ is retained as the best accuracy/training time ratio.

\begin{table}
	\begin{subtable}[h]{\textwidth}
\centering
		\footnotesize
		\caption{Model performance for various $(n,s)$ pairs}
		\label{mse_ns}
		\centering
		\begin{tabular}{cccc} 
																												\toprule
\textbf{Pairs $(n,s)$}		& \textbf{$ \overline{\mathcal{L}_{MSE}} $} 	&	\textbf{Training time (hours)}	&	\textbf{Patches per snapshot} 	\\\midrule
200,300			    	& \num{1.08e-6}						&	1.06	    					& 1							\\\midrule 
100,150				& \num{3.09e-7}						&	1.85						& 4							\\\midrule 
50,75				& \num{3.36e-7}						&	2.38						& 20  						\\\midrule 
20,30				& \num{1.77e-6}						&	3.34						& 120						\\\midrule 
10,15				& \num{1.94e-6}						&	7.09						& 480						\\\midrule 
6,9				    	& \num{1.98e-6}						&	62.26					& 1320						\\\midrule
2,3				    	& \num{7.85e-6}						&	128.5					& 12000						\\\bottomrule
		\end{tabular}
	\end{subtable}
	
	\medskip

	\begin{subtable}[h]{\textwidth}
\centering
		\footnotesize
		\caption{Model performance for varying stride $s$, with $n=50$}
		\label{mse_s}
		\centering
		\begin{tabular}{cccc} 
																												\toprule
\textbf{Pairs $(n,s)$}		& \textbf{$ \overline{\mathcal{L}_{MSE}} $}	&	\textbf{Training time (hours)}	&	\textbf{Patches per snapshot}  	\\\midrule
50,300			    	& \num{3.98e-7}						&	1.93						& 2							\\\midrule 	
50,150				& \num{4.06e-7}						&	2.34						& 6							\\\midrule 		
50,75				& \num{3.36e-7}						&	2.38						& 20							\\\midrule
50,30				& \num{3.57e-7}						&	10.2						& 99							\\\bottomrule

		\end{tabular}
	\end{subtable}
	
	\medskip

    	\begin{subtable}[h]{\textwidth}
\centering
		\footnotesize
		\caption{Model performance for varying batch size $b$, with $(n,s) = (50,75)$}
		\label{mse_b}
		\centering
		\begin{tabular}{ccc} 
																				\toprule
\textbf{Batch size}	& \textbf{$ \overline{\mathcal{L}_{MSE}} $}	&	\textbf{Training time (hours)}   \\\midrule
256             		& \num{3.10e-6}					        &	0.86						\\\midrule
128             		& \num{2.34e-6}					        &	1.07						\\\midrule
64			    	& \num{1.48e-6}					   	&	1.40						\\\midrule 	
32				& \num{3.36e-7}						&	2.38						\\\midrule 		
16				& \num{4.09e-7}						&	3.37						\\\midrule 			
8				& \num{2.85e-7}						&	9.80						\\\bottomrule 			
		\end{tabular}
	\end{subtable}
	
	\medskip
	\medskip

	\caption{\textbf{Model performance for varying $(n,s,b)$ parameters.} (\ref{mse_ns}) Model performance for various $(n,s)$ pairs, with the constraint $s=1.5 \times n$. Best validation performance is obtained for $(100,150)$, but the close performance of $(50,75)$ and its larger amount of generated snapshots make it a more versatile candidate. (\ref{mse_s}) Model performance for varying stride size $s$, with $n=50$. Best performance is obtained for $s=75$, although other stride values present closer performance levels. (\ref{mse_b}) Model performance for varying batch size $b$, with $(n,s) = (50,75)$. Although best performance was obtained for $b=8$, batch sizes of 16 and 32 made no significant different in validation error. Hence, faster training was privileged, and $b=32$ was retained.}
	\label{table:mse_parametric}
\end{table}

\section{Conclusions}

In this article, we have demonstrated the deployment of a robust deep learning model for predicting Spalart-Allmaras eddy viscosities. The method of patch-based training works by dividing the full-scale samples into patches, in order to let the model learn multiple local feature mappings, instead of learning monolithic full-scale features. Applied to an auto-encoder architecture, it was observed that patch-based training led to training and validation errors one order of magnitude lower than standard full-scale training, and was able able to efficiently learn local mappings from multiple datasets with different features, which was not the case of full-scale training method. For practical CFD purposes, a local patch-based model would be of great importance so that any input fluid domain, either full or in parts by region of interest, can be split into patches and passed to the model to predict the quantities of interest. Hence, patch-based training holds an important potential to improve the usability of trained models in the coupling with CFD solvers. \red{Deploying a trained model to solve for turbulent viscosity inside a CFD solver is regarded as a future extension of the present work.}

\section*{Acknowledgements} 
This work is supported by the Carnot M.I.N.E.S. Institute through the MINDS - Mines Initiative for Numerics and Data Science project.

\section{Data availability statement}
\label{section:open_source}

The source code associated with the current article is available on the following GitHub repository: \url{https://github.com/jviquerat/cnn_spallart_allmaras} \footnote{The code will be released upon publication of the present manuscript}.

\bibliographystyle{unsrt}
\bibliography{bibfile}

\begin{thebibliography}{10}

\bibitem{spalart1992one}
Philippe Spalart and Steven Allmaras.
\newblock A one-equation turbulence model for aerodynamic flows.
\newblock In {\em 30th aerospace sciences meeting and exhibit}, page 439, 1992.

\bibitem{ferziger2002computational}
Joel~H Ferziger, Milovan Peri{\'c}, and Robert~L Street.
\newblock {\em Computational methods for fluid dynamics}, volume~3.
\newblock Springer, 2002.

\bibitem{spalart2000strategies}
Philippe~R Spalart.
\newblock Strategies for turbulence modelling and simulations.
\newblock {\em International journal of heat and fluid flow}, 21(3):252--263,
  2000.

\bibitem{milano2002neural}
Michele Milano and Petros Koumoutsakos.
\newblock Neural network modeling for near wall turbulent flow.
\newblock {\em Journal of Computational Physics}, 182(1):1--26, 2002.

\bibitem{Yarlanki2012}
S.~Yarlanki, B.~Rajendran, and H.~Hamann.
\newblock Estimation of turbulence closure coefficients for data centers using
  machine learning algorithms.
\newblock In {\em 13th InterSociety Conference on Thermal and Thermomechanical
  Phenomena in Electronic Systems}, pages 38--42, May 2012.

\bibitem{CHEUNG2011}
Sai~Hung Cheung, Todd~A. Oliver, Ernesto~E. Prudencio, Serge Prudhomme, and
  Robert~D. Moser.
\newblock Bayesian uncertainty analysis with applications to turbulence
  modeling.
\newblock {\em Reliability Engineering System Safety}, 96(9):1137 -- 1149,
  2011.
\newblock Quantification of Margins and Uncertainties.

\bibitem{Kato2014}
Hiroshi Kato and Shigeru Obayashi.
\newblock {\em Data Assimilation for Turbulent Flows}, chapter AIAA SciTech
  Forum.
\newblock American Institute of Aeronautics and Astronautics, Jan 2014.
\newblock 0.

\bibitem{PARISH2016758}
Eric~J. Parish and Karthik Duraisamy.
\newblock A paradigm for data-driven predictive modeling using field inversion
  and machine learning.
\newblock {\em Journal of Computational Physics}, 305:758 -- 774, 2016.

\bibitem{xiao2016quantifying}
H~Xiao, J-L Wu, J-X Wang, R~Sun, and CJ~Roy.
\newblock Quantifying and reducing model-form uncertainties in
  reynolds-averaged navier--stokes simulations: A data-driven, physics-informed
  bayesian approach.
\newblock {\em Journal of Computational Physics}, 324:115--136, 2016.

\bibitem{wang2017physics}
Jian-Xun Wang, Jin-Long Wu, and Heng Xiao.
\newblock Physics-informed machine learning approach for reconstructing
  reynolds stress modeling discrepancies based on dns data.
\newblock {\em Physical Review Fluids}, 2(3):034603, 2017.

\bibitem{singh2017data}
Anand~Pratap Singh, Racheet Matai, Asitav Mishra, Karthikeyan Duraisamy, and
  Paul~A Durbin.
\newblock Data-driven augmentation of turbulence models for adverse pressure
  gradient flows.
\newblock In {\em 23rd AIAA Computational Fluid Dynamics Conference}, page
  3626, 2017.

\bibitem{singh2017machine}
Anand~Pratap Singh, Shivaji Medida, and Karthik Duraisamy.
\newblock Machine-learning-augmented predictive modeling of turbulent separated
  flows over airfoils.
\newblock {\em AIAA Journal}, pages 1--13, 2017.

\bibitem{singh2016using}
Anand~Pratap Singh and Karthik Duraisamy.
\newblock Using field inversion to quantify functional errors in turbulence
  closures.
\newblock {\em Physics of Fluids}, 28(4):045110, 2016.

\bibitem{singh2017augmentation}
Anand~Pratap Singh, Karthikeyan Duraisamy, and Ze~Jia Zhang.
\newblock Augmentation of turbulence models using field inversion and machine
  learning.
\newblock In {\em 55th AIAA Aerospace Sciences Meeting}, page 0993, 2017.

\bibitem{ling2015evaluation}
Julia Ling and J~Templeton.
\newblock {Evaluation of machine learning algorithms for prediction of regions
  of high Reynolds averaged Navier Stokes uncertainty}.
\newblock {\em Phys. Fluids}, 27(8):085103, 2015.

\bibitem{ling2016reynolds}
Julia Ling, Andrew Kurzawski, and Jeremy Templeton.
\newblock Reynolds averaged turbulence modelling using deep neural networks
  with embedded invariance.
\newblock {\em Journal of Fluid Mechanics}, 807:155--166, 2016.

\bibitem{vollant2017subgrid}
Antoine Vollant, Guillaume Balarac, and C~Corre.
\newblock Subgrid-scale scalar flux modelling based on optimal estimation
  theory and machine-learning procedures.
\newblock {\em J. Turbul.}, 18(9):854--878, 2017.

\bibitem{maulik2017neural}
Romit Maulik and Omer San.
\newblock A neural network approach for the blind deconvolution of turbulent
  flows.
\newblock {\em Journal of Fluid Mechanics}, 831:151--181, 2017.

\bibitem{fukami2019synthetic}
Kai Fukami, Yusuke Nabae, Ken Kawai, and Koji Fukagata.
\newblock Synthetic turbulent inflow generator using machine learning.
\newblock {\em Physical Review Fluids}, 4(6):064603, 2019.

\bibitem{mohan2019compressed}
Arvind Mohan, Don Daniel, Michael Chertkov, and Daniel Livescu.
\newblock Compressed convolutional lstm: An efficient deep learning framework
  to model high fidelity 3d turbulence.
\newblock {\em arXiv preprint arXiv:1903.00033}, 2019.

\bibitem{beck2019deep}
Andrea Beck, David Flad, and Claus-Dieter Munz.
\newblock Deep neural networks for data-driven les closure models.
\newblock {\em Journal of Computational Physics}, 398:108910, 2019.

\bibitem{kim2019deep}
Junhyuk Kim and Changhoon Lee.
\newblock Deep unsupervised learning of turbulence for inflow generation at
  various reynolds numbers.
\newblock {\em arXiv preprint arXiv:1908.10515}, 2019.

\bibitem{fukami2019super}
Kai Fukami, Koji Fukagata, and Kunihiko Taira.
\newblock Super-resolution reconstruction of turbulent flows with machine
  learning.
\newblock {\em J. Fluid Mech.}, 870:106--120, 2019.

\bibitem{fukami2020machine}
Kai Fukami, Koji Fukagata, and Kunihiko Taira.
\newblock Machine learning based spatio-temporal super resolution
  reconstruction of turbulent flows.
\newblock {\em arXiv preprint arXiv:2004.11566}, 2020.

\bibitem{zhao2019turbulence}
Yaomin Zhao, Harshal~D Akolekar, Jack Weatheritt, Vittorio Michelassi, and
  Richard~D Sandberg.
\newblock Turbulence model development using cfd-driven machine learning.
\newblock {\em arXiv preprint arXiv:1902.09075}, 2019.

\bibitem{taghizadeh2020turbulence}
Salar Taghizadeh, Freddie~D Witherden, and Sharath~S Girimaji.
\newblock Turbulence closure modeling with data-driven techniques: physical
  compatibility and consistency considerations.
\newblock {\em arXiv preprint arXiv:2004.03031}, 2020.

\bibitem{viquerat2020supervised}
Jonathan Viquerat and Elie Hachem.
\newblock A supervised neural network for drag prediction of arbitrary 2d
  shapes in laminar flows at low reynolds number.
\newblock {\em Computers \& Fluids}, page 104645, 2020.

\bibitem{chen2021twin}
Junfeng Chen, Jonathan Viquerat, Frederic Heymes, and Elie Hachem.
\newblock A twin-decoder structure for incompressible laminar flow
  reconstruction with uncertainty estimation around 2d obstacles.
\newblock {\em arXiv preprint arXiv:2104.03619}, 2021.

\bibitem{duraisamy2019turbulence}
Karthik Duraisamy, Gianluca Iaccarino, and Heng Xiao.
\newblock Turbulence modeling in the age of data.
\newblock {\em Annu. Rev. Fluid Mech.}, 51:357--377, 2019.

\bibitem{zhang2019recent}
Xinlei Zhang, Jinlong Wu, Olivier Coutier-Delgosha, and Heng Xiao.
\newblock Recent progress in augmenting turbulence models with physics-informed
  machine learning.
\newblock {\em Journal of Hydrodynamics}, 31(6):1153--1158, 2019.

\bibitem{tracey2015machine}
Brendan~D Tracey, Karthikeyan Duraisamy, and Juan~J Alonso.
\newblock A machine learning strategy to assist turbulence model development.
\newblock In {\em 53rd AIAA aerospace sciences meeting}, page 1287, 2015.

\bibitem{liang2019developing}
SUN Liang, AN~Wei, LIU Xuejun, and LYU Hongqiang.
\newblock On developing data-driven turbulence model for dg solution of rans.
\newblock {\em Chinese Journal of Aeronautics}, 32(8):1869--1884, 2019.

\bibitem{maulik2020turbulent}
Romit Maulik, Himanshu Sharma, Saumil Patel, Bethany Lusch, and Elise Jennings.
\newblock A turbulent eddy-viscosity surrogate modeling framework for
  reynolds-averaged navier-stokes simulations.
\newblock {\em Computers \& Fluids}, page 104777, 2020.

\bibitem{pal2020deep}
Anikesh Pal.
\newblock Deep learning emulation of subgrid-scale processes in turbulent shear
  flows.
\newblock {\em Geophysical Research Letters}, 47(12):e2020GL087005, 2020.

\bibitem{long2015fully}
Jonathan Long, Evan Shelhamer, and Trevor Darrell.
\newblock Fully convolutional networks for semantic segmentation.
\newblock In {\em Proceedings of the IEEE conference on computer vision and
  pattern recognition}, pages 3431--3440, 2015.

\bibitem{farabet2012learning}
Clement Farabet, Camille Couprie, Laurent Najman, and Yann LeCun.
\newblock Learning hierarchical features for scene labeling.
\newblock {\em IEEE transactions on pattern analysis and machine intelligence},
  35(8):1915--1929, 2012.

\bibitem{pinheiro2014recurrent}
Pedro Pinheiro and Ronan Collobert.
\newblock Recurrent convolutional neural networks for scene labeling.
\newblock In {\em International conference on machine learning}, pages 82--90.
  PMLR, 2014.

\bibitem{pope2001turbulent}
Stephen~B Pope.
\newblock Turbulent flows, 2001.

\bibitem{guiza2020anisotropic}
G~Guiza, A~Larcher, A~Goetz, L~Billon, P~Meliga, and Elie Hachem.
\newblock Anisotropic boundary layer mesh generation for reliable 3d unsteady
  rans simulations.
\newblock {\em Finite Elements in Analysis and Design}, 170:103345, 2020.

\bibitem{rumsey2010description}
Chris Rumsey, Brian Smith, and George Huang.
\newblock Description of a website resource for turbulence modeling
  verification and validation.
\newblock In {\em 40th Fluid Dynamics Conference and Exhibit}, page 4742, 2010.

\bibitem{allmaras2012modifications}
Steven~R Allmaras and Forrester~T Johnson.
\newblock Modifications and clarifications for the implementation of the
  spalart-allmaras turbulence model.
\newblock In {\em Seventh international conference on computational fluid
  dynamics (ICCFD7)}, pages 1--11, 2012.

\bibitem{hachem2013immersed}
Elie Hachem, Stephanie Feghali, Ramon Codina, and Thierry Coupez.
\newblock Immersed stress method for fluid--structure interaction using
  anisotropic mesh adaptation.
\newblock {\em International journal for numerical methods in engineering},
  94(9):805--825, 2013.

\bibitem{rodi1997status}
W~Rodi, JH~Ferziger, M~Breuer, and M~Pourquie.
\newblock Status of large eddy simulation: results of a workshop.
\newblock {\em Transactions-American Society of Mechanical Engineers Journal of
  Fluids Engineering}, 119:248--262, 1997.

\bibitem{chen2021graph}
Junfeng Chen, Elie Hachem, and Jonathan Viquerat.
\newblock Graph neural networks for laminar flow prediction around random 2d
  shapes.
\newblock {\em arXiv preprint arXiv:2107.11529}, 2021.

\bibitem{kingma2014adam}
Diederik~P Kingma and Jimmy Ba.
\newblock Adam: A method for stochastic optimization.
\newblock {\em arXiv preprint arXiv:1412.6980}, 2014.

\bibitem{abadi2016tensorflow}
Martin Abadi, Paul Barham, Jianmin Chen, Zhifeng Chen, Andy Davis, Jeffrey
  Dean, Matthieu Devin, Sanjay Ghemawat, Geoffrey Irving, Michael Isard, et~al.
\newblock Tensorflow: A system for large-scale machine learning.
\newblock In {\em 12th USENIX Symposium on Operating Systems Design and
  Implementation (OSDI 16)}, pages 265--283, 2016.

\bibitem{hinton2006reducing}
Geoffrey~E Hinton and Ruslan~R Salakhutdinov.
\newblock Reducing the dimensionality of data with neural networks.
\newblock {\em science}, 313(5786):504--507, 2006.

\bibitem{patil2019development}
Aakash~Vijay Patil and Corentin Lapyere.
\newblock Development of deep learning methods for inflow turbulence
  generation.
\newblock {\em arXiv preprint arXiv:1910.06810}, 2019.

\bibitem{nair2010rectified}
Vinod Nair and Geoffrey~E Hinton.
\newblock Rectified linear units improve restricted boltzmann machines.
\newblock In {\em Icml}, pages 285--319, 2010.

\end{thebibliography}

\end{document}